\PassOptionsToPackage{table}{xcolor}
\documentclass[sigconf, screen]{acmart}

\copyrightyear{2021}
\acmYear{2021}
\setcopyright{ccbysa}
\acmConference[ASIA CCS '21] {2021 ACM Asia Conference on Computer and Communications Security}{June 7--11, 2021}{Hong Kong, Hong Kong}
\acmBooktitle{2021 ACM Asia Conference on Computer and Communications Security (ASIA CCS '21), June 7--11, 2021, Hong Kong, Hong Kong}
\acmPrice{}
\acmISBN{978-1-4503-8287-8/21/06}
\acmDOI{10.1145/3433210.3453096}
\graphicspath{ {./figures/} }
\usepackage{xspace}
\usepackage{todonotes}
\usepackage{paralist}
\usepackage{xcolor}
\usepackage{balance}

\usepackage{pifont}
\newcommand{\xmark}{\ding{55}}
\newcommand{\cmark}{\ding{51}}%

\newcommand{\lavaset}{\textsf{lava\_set()}\xspace}
\newcommand{\lavaget}{\textsf{lava\_get()}\xspace}
\newcommand{\dataflow}{\textsf{dataflow}\xspace}

\newcommand{\vda}{\hat{A}_{12}}
\newcommand{\duktape}{\textsf{duktape}\xspace}
\newcommand{\sqliteB}{\textsf{sqliteB}\xspace}
\newcommand{\yamlBtwo}{\textsf{yamlB2}\xspace}
\newcommand{\pcreB}{\textsf{pcreB}\xspace}
\newcommand{\grepBtwo}{\textsf{grepB2}\xspace}
\newcommand{\newgrepS}{\textsf{newgrepS}\xspace}
\newcommand{\tcpdump}{\textsf{tcpdump}\xspace}
\newcommand{\tcpdumpB}{\textsf{tcpdumpB}\xspace}
\newcommand{\file}{\textsf{file}\xspace}
\newcommand{\fileBthree}{\textsf{fileB3}\xspace}
\newcommand{\fileSfour}{\textsf{fileS4}\xspace}

\newcommand{\tinyexprBtwo}{\textsf{tinyexprB2}\xspace}
\newcommand{\hpc}[1]{\emph{HPC-#1}\xspace}
\newcommand{\Rodeoday}{Rode\-0\-day\xspace}

\newcommand{\yentry}{Y}
\newcommand{\nentry}{N}

\begin{document}
\fancyhead{}

\title{Evaluating Synthetic Bugs}

\author{Joshua Bundt}
\affiliation{%
\institution{Northeastern University}
    \country{}}
\email{bundt.j@northeastern.edu}

\author{Andrew Fasano}
\thanks{\noindent \scriptsize{DISTRIBUTION STATEMENT A. Approved for public release. Distribution is unlimited.  This material is based upon work supported by the Department of Defense under Air Force Contract No. FA8702-15-D-0001. Any opinions, findings, conclusions or recommendations expressed in this material are those of the author(s) and do not necessarily reflect the views of the Department of Defense.}\endgraf
\noindent \scriptsize{Delivered to the U.S. Government with Unlimited Rights, as defined in DFARS Part 252.227-7013 or 7014 (Feb 2014). Notwithstanding any copyright notice, U.S. Government rights in this work are defined by DFARS 252.227-7013 or DFARS 252.227-7014 as detailed above. Use of this work other than as specifically authorized by the U.S. Government may violate any copyrights that exist in this work.}
}

\affiliation{%
\institution{Northeastern University}
    \country{}}
\affiliation{%
\institution{MIT Lincoln Laboratory}
    \country{}}
\email{fasano@mit.edu}

\author{Brendan Dolan-Gavitt}
\affiliation{%
\institution{New York University}
    \country{}}
\email{brendandg@nyu.edu}

\author{William Robertson}
\affiliation{%
\institution{Northeastern University}
    \country{}}
\email{w.robertson@northeastern.edu}

\author{Tim Leek}
\affiliation{%
\institution{MIT Lincoln Laboratory}
    \country{}}
\email{tleek@ll.mit.edu}

\date{}

\begin{abstract}
Fuzz testing has been used to find bugs in programs since the 1990s,
but despite decades of dedicated research, there is still no consensus
on which fuzzing techniques work best.
One reason for this is the paucity of ground truth: bugs in real programs
with known root causes and triggering inputs are difficult to collect at a meaningful scale.
Bug injection technologies that add synthetic bugs into real programs
seem to offer a solution, but the differences in finding
these synthetic bugs versus organic bugs have not previously been explored
at a large scale.
Using over 80 years of CPU time, we ran eight fuzzers across 20 targets
from the Rode0day bug-finding competition and the LAVA-M corpus.
Experiments were standardized with respect to compute resources and metrics gathered.
These experiments show differences in fuzzer performance as well as the impact of various configuration options.
For instance, it is clear that integrating symbolic execution with mutational fuzzing is very effective and that using dictionaries improves performance.
Other conclusions are less clear-cut;
for example, no one fuzzer beat all others on all tests.
It is noteworthy that no fuzzer found any organic bugs (i.e., one reported in a CVE), despite 50 such bugs being available for discovery in the fuzzing corpus.
A close analysis of results revealed a possible explanation: a dramatic difference between where synthetic and organic bugs live with respect to the ``main path'' discovered by fuzzers.
We find that recent updates to bug injection systems have made
synthetic bugs more difficult to discover,
but they are still significantly easier to find than organic bugs in our target programs.
Finally, this study identifies flaws in bug injection techniques and suggests a number of axes along which synthetic bugs should be improved.
\end{abstract}



\maketitle

\section{Introduction}
\label{section:introduction}
Fuzz testing, or fuzzing, is currently one of
the best techniques for vulnerability discovery.
Since its introduction in 1990~\cite{miller_1990_empiricalstudyreliability},
a variety of fuzzing techniques have been explored including
grammar-based fuzzing~\cite{godefroid2008grammar};
fuzzing combined with symbolic execution and SMT solvers~\cite{stephens_2016_drilleraugmentingfuzzing};
taint-based fuzzing~\cite{ganesh_2009};
and fuzzing with neural networks~\cite{she_2019_neuzzefficientfuzzing}.
Today, fuzzing is used widely and at scale in industry (e.g., Sage~\cite{godefroid_2008_automatedwhiteboxfuzz}, ClusterFuzz~\cite{google_2020_clusterfuzz}, and OSS-Fuzz~\cite{google_2020_ossfuzz}).
We refer interested readers to a recent survey on fuzzing techniques for a comprehensive overview of the field~\cite{manes_2019_artscienceengineering}.

Automated vulnerability discovery is essential to both software developers interested in deploying secure software as well as hackers interested in exploiting software.
As such, there is a clear need to understand what vulnerability discovery techniques actually work in practice and in which contexts. 
Furthermore, accurately quantifying and describing the performance of novel approaches to vulnerability discovery tools and techniques is necessary to advance the state of the art.

Fortunately, significant prior work has laid out guidelines and exposed pitfalls concerning how fuzzers should be evaluated~\cite{klees_2018_evaluatingfuzztesting}.
A critical component of fuzzer evaluations is the ``bug corpus,'' which has historically been a combination of previously-discovered bugs and new (0-day) discoveries attributed to the technique under evaluation.
In 2016, LAVA introduced synthetic bug generation, in part, to overcome the limitations of relying on known vulnerabilities for fuzzer evaluation~\cite{dolan-gavitt_2016_lavalargescaleautomated}.

In the years since the release of LAVA and its most well-known benchmark corpus, LAVA-M, a large body of work has used synthetic bug injection in their evaluations.
However, it is unclear that LAVA bugs are representative of organic (i.e., non-synthetic) bugs.
Since LAVA bugs are commonly found by modern fuzzers while organic bugs remain undiscovered,
perhaps these synthetic bugs are easier to find than non-synthetic bugs, which are organically (and unintentionally) added into programs.
If there are significant differences between the methods of discovering LAVA bugs vs.~discovering organic bugs, it would call into question the assumption that a fuzzer capable of efficiently finding LAVA-injected bugs will perform similarly with organic bugs.

We performed a large-scale measurement study to validate or refute the conventional wisdom regarding the utility of synthetic bug generation. We evaluated eight distinct fuzzers on 20 targets over the course of eight experiments. These experiments test the selected fuzzers' ability to discover bugs injected by LAVA in its default configuration, alternative LAVA configurations (some of which we implemented), as well as bugs injected by other synthetic bug-injection systems, and by hand.
In total, we ran these fuzzers for 733K CPU-hours, or just over 83.5 CPU-years.

Our evaluation reveals five key findings:
\begin{enumerate}
\item Symbolic execution integrated with mutational fuzzing is highly effective.
\item Gray-box, coverage-guided fuzzers can effectively find LAVA bugs with simple techniques such as dictionaries or comparison splitting.
\item Injected bugs are biased towards a target program's \emph{main path} which skews analysis results.
\item Recent updates to LAVA increase the difficulty of discovering injected bugs, but synthetic bugs still differ significantly from organic bugs.
\item LAVA-M and a portion of the evaluated challenges exhibit fundamental weaknesses that should preclude them from future fuzzer evaluations.
\end{enumerate}

Our experiments were not able to reproduce bugs discovered in prior work despite investing similar resources, providing further evidence that organic bugs are more difficult to find than LAVA bugs.
Based on these findings, we identify several promising directions for improving synthetic bug injection to more closely model the difficulty of organic bug discovery.
The first is requiring attacker-controlled data to satisfy constraints beyond simple equality checks.
Injection techniques should also aim to place bugs ``far away'' from commonly executed code.
These techniques should be resistant to bug-finding based on dictionary extraction or comparison splitting.
Finally, bug injection approaches should better measure the path constraint solving capabilities of hybrid fuzzers.
In support of enabling scientific replication of our results, we have published our tools and data at \url{https://rode0day.gitlab.io/evaluation}.

The remainder of this paper is structured as follows.
In~\S\ref{section:background}, we present background information on fuzzing and synthetic bug generation.
\S\ref{section:methodology} presents our measurement methodology, and \S\ref{section:setup} outlines the experiments we conducted.
In~\S\ref{section:evaluation}, we present our findings and key takeaways from the measurement.
We discuss implications of these findings in~\S\ref{section:future}, and conclude the paper in~\S\ref{section:conclusions}.

\section{Background and Related Work}
\label{section:background}

Due to its effectiveness and efficiency, fuzzing is the favored technique for automated vulnerability discovery.
At a high level, a fuzzer prepares inputs, or test cases, to a program under test, observes the program's behavior as it executes over these test cases, and records security violations for later examination by an analyst.
White-box fuzzers use information derived from the program to generate test cases, for instance via symbolic execution~\cite{godefroid_2007_randomtestingsecurity,godefroid_2008_automatedwhiteboxfuzz}, while grey-box fuzzers~\cite{zalewski_2020_americanfuzzylop,google_2020_libfuzzerlibrarycoverageguided,rawat_2017_vuzzerapplicationawareevolutionary} use partial information such as coverage for the same purpose.  
Black-box fuzzers~\cite{aitel_2002_introductionspikefuzzer,peachtech_2020_peachfuzzer,hocevar_2020_samhocevarzzuf,cmusei_2020_certcccertfuzz}, on the other hand, eschew program analysis for faster input generation and thus trade off this insight into the program state space for simplicity and speed.  On each execution of a test case, fuzzers typically record information that influence the scheduling~\cite{cmusei_2020_certcccertfuzz,woo_2013_schedulingblackboxmutational,bohme_2016_coveragebasedgreyboxfuzzing} or generation of subsequent test cases. For white-box fuzzers, this could be path constraints over test case bytes, while for grey-box fuzzers this could be coverage metrics in terms of blocks or edges.  Test case generation itself can range from purely random mutation~\cite{miller_1990_empiricalstudyreliability,google_2020_googlehonggfuzz,zalewski_2020_americanfuzzylop} to model-guided~\cite{aitel_2002_introductionspikefuzzer,peachtech_2020_peachfuzzer,holler_2012_fuzzingcodefragments,peng_2018_tfuzzfuzzingprogram} input generation parameterized by information derived from prior execution traces.  Crashes are typically used as behavioral evidence of a security violation, and specialized instrumentation such as AddressSanitizer (ASan)~\cite{serebryany_2012_addresssanitizerfastaddress} can be used to force subtle run-time security violations to produce crashes when they otherwise would not.

\subsection{Evaluating Fuzzers}
\label{section:evaluating_fuzzers}
Fuzzers are typically evaluated on their ability to generate inputs over some period of time that increase coverage of target programs, discover bugs, or a combination of both.
Coverage can be defined in multiple ways; block, edge, and source code coverage is common.
Although covering vulnerable code is a prerequisite to bug discovery, the goal of fuzzing is to find bugs, not solely to increase coverage.
With this goal in mind, fuzzers are often evaluated on their ability to find known bugs (n-days) or unknown bugs (0-days).

Unfortunately, fuzzer evaluations are severely limited by a small supply of known bugs.
Public bug reports and CVE entries often lack sufficient detail to determine if a fuzzer has rediscovered a given bug.
Some high-quality, manually-curated bug corpora (e.g., Magma~\cite{hazimeh2020magma}),
aim to rectify this, but these are still fairly small.
Furthermore, using known bugs to evaluate fuzzers may bias evaluations towards incremental improvements on the approaches that previously discovered those bugs.

Automated synthetic bug-injection frameworks such as LAVA~\cite{dolan-gavitt_2016_lavalargescaleautomated}, Apocalypse~\cite{roy2018bug}, and EvilCoder~\cite{pewny2016evilcoder} provide an alternative path to evaluating fuzzers.
These frameworks automatically insert a large number of synthetically-generated bugs into existing programs which can then be used to evaluate fuzzers.
When a fuzzer fails to find any new vulnerabilities in an application with no known bugs, it can only be evaluated in terms of coverage.
However, if a program has known synthetic bugs, fuzzers can be evaluated on their ability to find these bugs.

\subsection{LAVA-based Bug Injection}
\begin{figure*}[t]
    \includegraphics[width=0.9\textwidth]{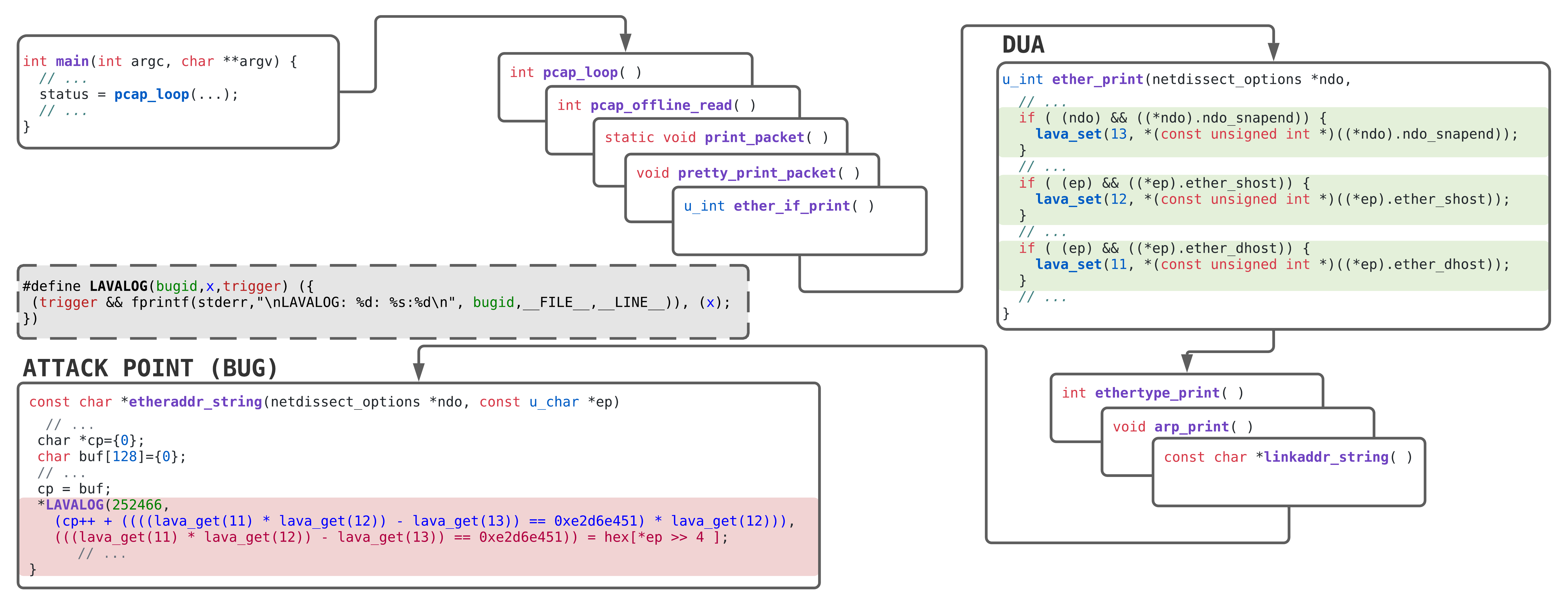}
    \caption{\tcpdumpB bug 252466. {\mdseries Example of a multi-DUA, \lavaset bug. Dead, unused, and available (DUA) bytes of the input are stored in global variables (highlighted in green).  Later, these values are retrieved and conditionally trigger a crashing bug (highlighted in red).}}
    \label{figure:lavabug_252466}
    \Description{Block diagram of LAVA bug 252466 from tcpdumpB}
\end{figure*}

Of the three synthetic bug-injection systems mentioned above, LAVA was the only one to release publicly available corpora of buggy programs: LAVA-M and LAVA-1.
The LAVA-M corpus is commonly used for fuzzing evaluation in the literature~\cite{rode0day_ieee}.
LAVA builds off of the PANDA~\cite{dolan2015repeatable} whole-system dynamic analysis platform to conduct a dynamic taint analysis on input files ingested by target applications.
The system begins with preprocessed C source code and uses source code rewriting and a dynamic taint analysis to identify
attacker-controlled
data referred to as DUAs\footnote{Dead (not used in many branches), Uncomplicated (not a complex function of input bytes), and Available at this point in the program trace} that can be modified without affecting the behavior of the target application.
During its analysis, LAVA also identifies source locations where bugs can be injected.
After the analysis concludes, LAVA injects memory corruption bugs at these locations which are triggered depending on a DUA value.
The modified program is then tested to ensure the bugs can be triggered and that benign inputs do not cause crashes.
In the LAVA-M/LAVA-1 data set, each bug triggered is a simple comparison of a constant value to a single DUA value.
Each DUA is assigned and retrieved using the \lavaset and \lavaget functions respectively which store and load DUAs from a global array.

To address concerns about LAVA bug realism,
Sridhar~\cite{sridhar2018adding} updated LAVA to support ``\dataflow mode'' where a local array of DUAs is passed by reference between functions.
Inspired by this work, we extended LAVA to support a ``\emph{multi-DUA} mode'' where injected bugs are triggered by one of the following expressions relating
three DUAs ($x, y, z$) and a per-bug random constant ($C$):
\begin{align*}
    C(x + y) &= z \\
    xy - z &= C \\ 
    (x + 2)(y + 3)(z + 1) &= C
\end{align*}
If the expression is satisfied, the DUA values will control how memory is corrupted.
An example multi-DUA bug 
is shown in Fig.~\ref{figure:lavabug_252466}.

LAVA also has limited support for \emph{coverage bugs} where unconditional bugs are injected on a coverage frontier.\footnote{\url{https://github.com/panda-re/lava/tree/covbugs/covbugs}}
These bugs are injected after analyzing a corpus of inputs (e.g., those from a fuzzing campaign) and combining the coverage from the provided inputs.
Unlike traditional LAVA bugs, coverage bugs are not accompanied with crashing inputs and may be impossible to trigger.

\subsection{Limitations of Synthetic Bug Injection}

Although there are strong arguments in favor of automated synthetic bug injection, existing work is known to suffer from several important shortcomings.  (We investigate the effect of these limitations on bug discoverability and fuzzer evaluations in~\S\ref{section:evaluation}.)

\emph{Injection coverage vs.~reachability trade-off.}
Systems that inject bugs along a path recorded using dynamic analysis and a concrete input are unable to inject bugs into uncovered code.  On the other hand, systems that inject bugs at arbitrary program points without concretely evaluating the bug are unable to determine if injected bugs can actually be triggered.

\emph{Limited bug types.}
Existing bug injection systems support a limited number of bug types (e.g., out-of-bounds array indexing) and adding new types is non-trivial.  While this shortcoming is not fundamental to the approach, it does practically limit the degree of insight that can be gained into fuzzers using synthetic bugs.

\emph{Bug realism and over-fitting.}
Injected bugs do not necessarily appear similar to organic bugs authored by humans.  This ``realism gap'' can take several forms.  For instance, in source code, injected bugs might use variable names that are clearly auto-generated or that are in some other way atypical of human naming.  At a semantic level, injected bugs might introduce control or data flows that are incongruent with the rest of the program.  This gap in turn opens the door for fuzzers to ``optimize for the benchmark,'' which can frustrate efforts to improve real-world performance.

To mitigate these shortcomings, fuzzer evaluations typically report detection performance on both synthetic benchmarks such as LAVA-M as well as a test set composed of real programs.  Nevertheless, use of synthetic benchmarks is eliciting increasing criticism from the security community, raising the question: \emph{Is synthetic bug generation an ecologically valid approach to fuzzer evaluation?}  We endeavor to answer this question in the remainder of this paper.

\section{Test Corpora and Methodology}
\label{section:methodology}

\begin{table*}[t]
\small
\centering
    \caption[caption]{Rode0day challenge programs selected for evaluation.}
    \label{table:challenges}
  \begin{tabular}{ l r r r r r c r c r r l } \toprule
      \textbf{Challenge} & \textbf{BT\textsuperscript{1}} & \textbf{BF\textsuperscript{1}} & \textbf{BI\textsuperscript{2}} & \textbf{S-DUA\textsuperscript{3}} & \textbf{M-DUA\textsuperscript{3}} & \textbf{Date} & \textbf{Version} & \textbf{CVEs} & \textbf{POCs} &  \textbf{SLOC} & \textbf{Description}\\ \midrule
      duktape    & 17  & 15  & L & 17  &    & 2018.11 & v2.3.0     &    &      & 60K    & JavaScript interpreter \\ \hline
      fileB3     & 72  & 72  & D & 31  & 31 & 2019.02 & v5.35      & 4  & 4   & 16K    & File type indentifier \\ \cline{1-6}
      fileS3     & 133 & 132 & L & 103 & 31 & 2019.09 &            &    &      &        \\ \cline{1-6}
      fileS4     & 103 & 103 & L & 77  & 26 & 2019.10 &            &    &      &        \\ \hline
      grepB2     & 45  & 45  & L & 31  & 14 & 2019.09 & v3.1       &    &      & 101K   & Pattern matcher \\ \hline
      jpegS3     & 97  & 1   & C &     &    & 2019.07 & v9c        &    &      & 29K    & JPEG image decoder \\ \hline 
      jqB        & 33  & 33  & D & 30  & 3  & 2019.01 & v1.6       &    &      & 40K    & JSON parser \\ \cline{1-6}
      jqB2       & 137 & 137 & D & 135 & 2  & 2019.03 &            &    &      &        \\ \cline{1-6}
      jqS3       & 29  & 28  & D & 26  & 3  & 2019.07 &            &    &      &        \\ \cline{1-6}
      jqS4       & 21  & 21  & L & 21  &    & 2019.09 &            &    &      &        \\ \hline
      newgrepS   & 4   & 4   & A &     &    & 2018.10 & v2.16      &    &      &        & Pattern matcher \\ \hline
      pcreB      & 106 & 106 & D & 73  & 33 & 2018.10 & v10.33-RC1 &    &      & 84K    & Regex library \\ \hline
      sqliteB    & 56  & 24  & L & 56  &    & 2019.05 & v3.29.0    & 18 & 14  & 160K   & SQL database \\ \hline
      tcpdumpB   & 76  & 59  & L & 74  & 2  & 2019.06 & v4.9.2     & 28 &  17 & 129K   & Packet analyzer \\ \hline
      tinyexprB2 & 4   & 3   & M &     &    & 2019.07 & master     &    &      & 682    & Math expression parser \\ \hline
      yamlB2     & 45  & 45  & L & 45  &    & 2019.07 & v0.1.7     &    &      & 10K    & YAML parser \\ \bottomrule
      \multicolumn{11}{l}{\footnotesize \textsuperscript{1}\textbf{BT} is total injected bugs. \textbf{BF} is total bugs found in all experiments. }\\
      \multicolumn{11}{l}{\footnotesize \textsuperscript{2}\textbf{BI} is bug injection type: data-flow (\textbf{D}); lava\_get (\textbf{L}); manual (\textbf{M}); coverage (\textbf{C}); Apocalypse (\textbf{A}).}\\
      \multicolumn{11}{l}{\footnotesize \textsuperscript{3}\#DUAs/bug: single-DUA (\textbf{S-DUA}); multi-DUA (\textbf{M-DUA}). }\\
  \end{tabular}
\end{table*}

\begin{figure*}[t]
    \centering
    \includegraphics[width=0.9\textwidth]{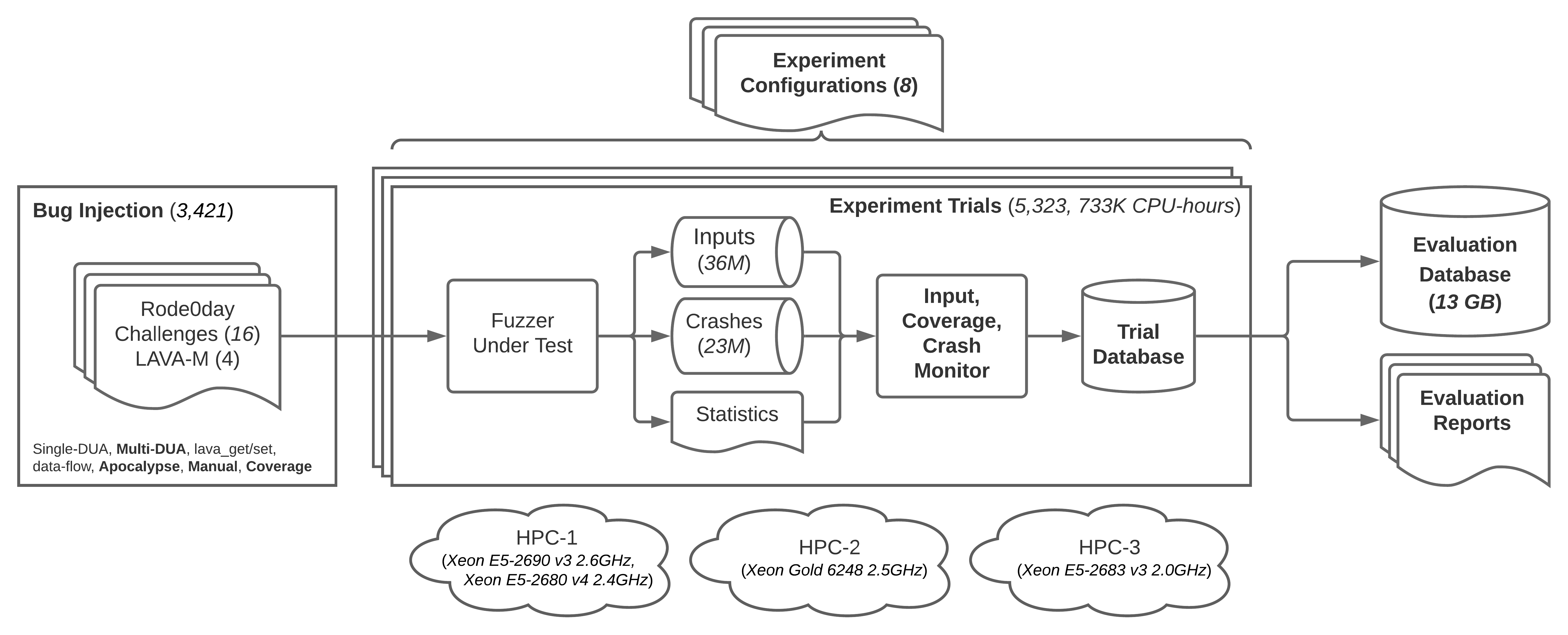}
    \caption{Overview of the experimental setup. {\mdseries Components in bold were developed, contributed, or collected by the authors.}}
    \label{figure:experiment}
    \Description{}
\end{figure*}

The primary aim of this work is to answer the question of whether synthetic bug injection is an ecologically valid approach to fuzzer evaluation---that is, whether conclusions drawn from a fuzzer's synthetic bug discovery performance can be reliably generalized to a real-world setting.  To answer this question, we devised a comprehensive experimental methodology to conduct experiments in support of this inquiry shown in Fig.~\ref{figure:experiment}.  In this section, we outline the challenge corpus, the fuzzers under test, and the nature of the data we collected to support the evaluation described in Sec.~\ref{section:evaluation}.

\subsection{Challenge Corpus}
\label{section:methodology:challenges}

Conducting an empirical evaluation of the utility of synthetic bugs for fuzzing evaluations requires obtaining a data set of challenges injected with synthetic bugs. We define a \emph{challenge} as a software artifact that has been injected with bugs; one original artifact can be injected multiple times to produce distinct challenges.

Similar to prior work~\cite{hazimeh2020magma}, we seek a data set consisting of a diverse set of challenges in terms of functionality, the type of input, and code complexity.  In addition, the challenges should be amenable to fuzz testing by popular fuzzers.  Finally, the injection procedure(s) should represent the current state of the art.

Luckily, the \Rodeoday~corpus satisfies each of these criteria. \Rodeoday~\cite{rode0day_ieee} is a continuous bug-finding competition that uses synthetic bugs.  The \Rodeoday~corpus\footnote{\url{https://rode0day.mit.edu/archive}} is a collection of 55 challenges built from 12 programs deployed in previous competitions.  Synthetic bugs were injected into these challenges primarily using LAVA~\cite{dolan-gavitt_2016_lavalargescaleautomated} in various configurations, although the corpus also contains a challenge generated by
Apocalypse~\cite{roy2018bug} and several challenges with manually-injected bugs.

In order to distill a feasible set of challenges from the \Rodeoday~corpus, we conducted a short evaluation of all 55 challenges with a representative subset of available fuzzers. During this challenge evaluation, we ensured that each fuzzer's compiler instrumentation toolchain could successfully compile the target, that the compiled challenge executed without error on the initial seed(s), and that the solution inputs for each injected bug caused the binary to crash as intended.  Finally, we ran the subset of fuzzers for a 15 minute fuzzing campaign with the default options to ensure that each fuzzer would successfully generate new, mutated inputs.
From this pool of feasible challenges, we selected a final set of evaluation challenges that varied along the following dimensions:

\begin{enumerate}
    \item bug injection method (distinct LAVA configurations, Apocalypse, or manual);
    \item program functionality and input type;
    \item percentage of bugs found during the Rode0day competition;
    \item type of LAVA bug trigger (\lavaset vs.~\dataflow); and,
    \item number of DUAs per LAVA bug (multi- vs.~single-DUA).
\end{enumerate}

The final challenge set was selected with the competing priorities of providing overall diversity while also supporting comparisons of specific aspects of synthetic bug injection.  In total, 16 challenges from 10 different open-source programs across 10 \Rodeoday~competitions comprise this set.
As these challenges are based on slightly outdated programs, 
limited number of CVEs describing organic bugs in these challenges are available.
An overview of the final Rode0day challenge set is shown in Table~\ref{table:challenges}.

\subsection{Fuzzer Test Corpus}
\label{section:methodology:fuzzers}

An overview of the eight mutational fuzzers selected for evaluation in this paper are shown in Table~\ref{table:fuzzers}.  These fuzzers were selected for evaluation using the following criteria.

\emph{State-of-the-art.}
Fuzzers that are widely regarded as representative of the state-of-the-art in mutational fuzzing were considered for inclusion in the evaluation.  This determination was made on the basis of published work and frequent inclusion in prior fuzzing evaluations.

\emph{Publicly available.} 
Unfortunately, some fuzzers do not release their source code, which hinders replication and validation of published results.  We chose to reduce the risk of unexplainable experimental anomalies by only considering fuzzers with source code available.

\emph{Testbed compatibility.} 
The fuzzer must be compatible with the constraints of the HPC environment used in the evaluation (described in Sec.~\ref{section:setup:infrastructure}). Unfortunately, this excluded some fuzzers from consideration in this work. For instance, REDQUEEN~\cite{aschermann2019redqueen} reported excellent performance on the LAVA-M data set.  However, that fuzzer requires Intel Processor Trace support and root privileges, neither of which were available in our evaluation environment.

\emph{Challenge corpus compatibility.}
The advantages of the fuzzer must be generally applicable to \Rodeoday~target challenges, or else it was excluded from consideration. For instance, AFLNET is a network protocol fuzzer which does not apply to any of the selected \Rodeoday~challenges.

\emph{Minimal preparation overhead.} 
Some fuzzers focus on a specific domain of inputs or program functionality.  For instance, AFL-smart requires a Peach fuzzer definition (i.e., a grammar describing the input language) for each challenge and was thus excluded. 

\emph{Diversity of implementation.} 
The fuzzer corpus should manifest diversity of implementation. In particular, although evaluating several AFL derivatives is almost unavoidable due to the popularity of AFL within the research community, we made an effort to include fuzzers without direct AFL lineage.

\begin{table}
\small
\centering
    \caption{Fuzzers evaluated.}\label{table:fuzzers}
    \rowcolors{2}{gray!25}{white}
    \begin{tabular}{llccc} \toprule
        \textbf{Fuzzer} & \textbf{Version} & \textbf{Year} & \textbf{Binary} & \textbf{Source} \\ \midrule
        AFL~\cite{zalewski_2020_americanfuzzylop}    & v2.56b    & 2016 & \yentry & \yentry \\
        AFLplusplus~\cite{fiorald2020afl++} (AFL++)  & v2.62c    & 2020 & \yentry & \yentry \\
        FairFuzz~\cite{lemieux2018fairfuzz} (AFL-rb) & v2.52     & 2017 & \yentry & \yentry \\
        Angora~\cite{chen2018angora}                 & v1.2.2    & 2018 & \nentry & \yentry \\
        Eclipser~\cite{choi2019eclipser}             & v1.0      & 2019 & \yentry & \nentry \\
        Ankou~\cite{manes2020ankou}                  & v1.0      & 2020 & \yentry & \yentry \\
        Honggfuzz~\cite{swiecki2016honggfuzz}        & v2.1      & --   & \yentry & \yentry \\
        QSYM~\cite{yun_2018_qsympracticalconcolic}   & \#89a761d & 2018 & \yentry & \yentry \\ \bottomrule
    \end{tabular}
\end{table}

\subsection{Coverage Measurement}
\label{section:methodology:coverage}

Collecting unbiased coverage metrics during a fuzzing experiment is a non-trivial and under-discussed task. 
Previous research varies wildly in the metrics collected, the collection methods used, and the granularity of the metrics themselves. 
Many studies prefer to report the AFL statistic of paths, others report basic block or edge coverage, and some report source code line coverage. 
Empirical research has shown the pitfalls of poor coverage measures~\cite{simon2020aflcc}, but even when block or edge coverage is used, rarely is the method for measuring coverage reported.

For these reasons, we adopted the binary coverage tool used to evaluate REDQUEEN~\cite{aflq-fast-cov}. 
We extended the QEMU-based coverage tool to execute concurrently with the fuzzer, capturing coverage and statistics in real time.
We found that methods that measured coverage post-experiment suffered a loss of fidelity due to differences in fuzzer queue management behavior.
By running the monitor concurrently with the fuzzer and utilizing the Linux inotify API, we collected accurate timestamps when every edge or bug was discovered. 
Additionally, by utilizing the same set of binaries to record coverage for all experiments, there is no discrepancy in coverage and bug reporting between different compiled versions of the same source program.

Our monitor logged block coverage, edge coverage, and triggered bugs in real time to a SQLite database that recorded test cases and crash inputs as the files were written to the file-system. 
After the experiments were complete all experiment databases were merged into one relational database for analysis.

\subsection{Bug Reporting}
By utilizing a synthetic bug corpus, our experiments use ground truth in found bug accounting.
Without synthetic bugs, researchers must rely on a method of crash triaging to determine unique bug counts.
As Klees et al.~\cite{klees_2018_evaluatingfuzztesting} pointed out, even the best available heuristics (e.g., stack hashing) perform quite poorly in determining unique bugs.
In addition to unique bug counts, some research also reports the AFL metric of unique crashes which can suffer from bias and over-counting of distinct source code flaws.

\section{Experimental Setup}
\label{section:setup}

To answer research questions about the utility of synthetic bugs, we conducted eight distinct experiments
with various fuzzer configurations, computational resources, time resources, and target types.
To conduct these experiments at the required scale, we developed a fuzzer-agnostic orchestration framework
to run on High Performance Computing (HPC) environments.
Here, we describe those experiments and the infrastructure we created.

\subsection{Fuzzing Experiments}
\label{section:setup:experiments}

\begin{table}
\small
\centering
    \caption{Experiment parameters.}\label{table:experiments}
    \rowcolors{2}{gray!25}{white}
    \begin{tabular}{llrcrrlrl} \toprule
        \textbf{ID} & \textbf{Desc} & \textbf{Hrs} & \textbf{D} & \textbf{Tr} & \textbf{Fz} & \textbf{Tg} & \textbf{CPU/h} & \textbf{HPC} \\ \midrule
        e1 & Default options & 24  & \yentry{} & 25 & 5 & R     & 96K  & 1 \\
        e2 & No dictionary   & 24  & \nentry{} & 5  & 8 & R     & 35K  & 1,2 \\
        e3 & Binary only     & 24  & \nentry{} & 5  & 4 & R     & 15K  & 1  \\
        e4 & 64-bit binaries & 24  & \yentry{} & 5  & 6 & R     & 23K  & 1  \\
        e5 & LAVA-M          & 24  & \yentry{} & 5  & 8 & L     & 77K  & 1  \\
        e6 & x4 threads/CPUs & 24  & \yentry{} & 10 & 5 & R     & 8K   & 1  \\
        e7 & 7 days          & 168 & \yentry{} & 9  & 6 & R$^*$ & 218K & 3  \\ 
        e8 & 28 days         & 672 & \yentry{} & 3  & 5 & R$^*$ & 262K & 3  \\ \midrule 
        \rowcolor{white}
        \multicolumn{3}{l}{\textbf{Totals}} &  & 5,323 &  &     & 733K &    \\ \bottomrule 
        \multicolumn{9}{l}{\footnotesize Default options: 32-bit binaries, w/dictionary, x2 threads/CPUs, w/source}\\
        \rowcolor{white}
        \multicolumn{9}{l}{\footnotesize Columns: \textbf{D}: fuzzing dictionary, \textbf{Tr}: \# of trials, \textbf{Fz}: \# of fuzzers, \textbf{Tg}: targets fuzzed}\\
        \multicolumn{9}{l}{\footnotesize Targets fuzzed: R: Rode0day challenges, R$^*$: R minus some duplicates, L: LAVA-M.}\\
    \end{tabular}
\end{table}

Table~\ref{table:experiments} shows the eight experiments we designed to explore parameter trade-offs
and to reduce the potential for accidentally introducing biases into the evaluation.
Due to individual fuzzer limitations, not every fuzzer was run in every experiment.
As a resource optimization, some of the challenges based on the same original program were limited to just one version for long-running experiments. 
Details on these exclusions are presented in Appendix~\ref{appendix:exclusion}.

\subsubsection{Default Options}
The default options for our experiments follow recommendations from previous research~\cite{klees_2018_evaluatingfuzztesting} where possible. 
Fuzzers were launched with two CPUs/threads except where otherwise specified.  Two additional CPUs were allocated for our real-time analysis process.
Fuzzers designed to run in a hybrid mode such as QSYM and Angora were run with one QSYM/Angora process synchronizing with one AFL instance in accordance with author recommendations. 
All AFL instances were run with the \texttt{-S} option which prevents deterministic mode fuzzing and enables AFL to sync progress with another fuzzer or itself. 
Eclipser was the only fuzzer which did not support specifying multi-threaded execution or multi-process synchronization at the time of writing.  
We used the same seed input files, execution timeouts, and compiler options for each target across all experiments as described below.

\emph{Seed input files.} 
We used the seeds provided during the Rode0day competition for nearly all challenges as they were well-formed input files.
The only exception was for \tcpdump as the provided seed violated Angora's input size restriction.
To resolve this, an alternative seed was selected and used for all fuzzers.

\emph{Execution timeouts.} 
The creators of Rode0day provided recommended timeout values for slow targets which we used.
When no such recommendation was provided, the default timeout of each fuzzer was used.

\emph{Compiler options.}
Targets were compiled as 32-bit binaries with default options.
They were also compiled with coverage sanitizers as required by the various fuzzers,
but without memory or undefined behavior sanitizers since LAVA-injected bugs cause segmentation fault-induced crashes without the need for sanitizers.
Angora lacks support for 32-bit binaries, so it was evaluated across all experiments on 64-bit versions of the challenges.

\subsubsection{Parameters Evaluated}
Across the experiments, we evaluated how dictionary availability, source code availability, and architecture affect fuzzer performance.

\emph{Dictionary availability.}
A fuzzing dictionary is a set of tokens that can be used to mutate an input. 
AFL added dictionary support in version 0.96b (January 2015).\footnote{\url{https://lcamtuf.blogspot.com/2015/01/afl-fuzz-making-up-grammar-with.html}} 
The source code release for AFL contains dictionaries for many common file formats and others have created large collections of dictionaries.\footnote{\url{https://github.com/google/fuzzing/tree/master/dictionaries}} 
libFuzzer and Honggfuzz support AFL-style dictionaries~\cite{google_2020_libfuzzerlibrarycoverageguided}, and OSS-Fuzz integrates dictionaries into many of its fuzzing projects. 
Despite nearly universal support for this feature and the prevailing belief that fuzzing dictionaries can have a dramatic positive effect on fuzzer efficiency, most fuzzing papers do not report whether or not a dictionary was used in experiments or if one is even available.\footnote{One exception is ProFuzzer~\cite{you2019profuzzer} which attempts to identify the location of fields and their types within the program input.}

Fuzzing dictionaries are designed to increase coverage, but for single-DUA LAVA-injected bugs (which compare input against a 4-byte magic value), a dictionary of constants parsed from a disassembly of the fuzzing target can be very effective in improving bug finding.
This method of extracting constants from a compiled binary was reported by the LAVA authors, but was never used in a LAVA-M evaluation.\footnote{\url{https://moyix.blogspot.com/2016/07/fuzzing-with-afl-is-an-art.html}}
We conducted experiments with and without~(e2) a dictionary of constants extracted from the challenge binary to measure its impact on bug finding and coverage.

\emph{Binary vs.~source.}
Many of the \Rodeoday~challenges were deployed in the competitions as binary-only challenges, meaning that fuzzers that rely on source code instrumentation or analysis were not able to compete on those challenges.
The source code for those binary-only challenges was subsequently released after the competitions.  In order to evaluate more fuzzers, we chose to only run one set of experiments with binary-only challenges~(e3).

\emph{32-bit vs.~64-bit.}
LAVA is currently only capable of injecting bugs within a 32-bit binary.
Most \Rodeoday~challenges were deployed as 32-bit challenges, with a small handful deployed as 64-bit binaries.
Due to the preprocessing that occurs before LAVA bugs are injected, compiling LAVA challenges for 64-bit architectures is not always a straightforward task.
To support our evaluation, we manually patched the \Rodeoday~challenges used in this evaluation so that they would compile for 32-bit or 64-bit architectures.
Additionally, Angora \emph{only} operates on 64-bit binaries, so to provide a unbiased comparison, we conducted one experiment with 64-bit binaries for all fuzzers~(e4).

\subsection{Fuzzing Infrastructure}
\label{section:setup:infrastructure}


\begin{table}
\small
\setlength\tabcolsep{3.5pt} 
\centering
    \caption{HPC specifications.}
    \label{table:hpcs}
    \begin{tabular}{llllcl} \toprule
        \textbf{HPC} & \textbf{Host OS} & \textbf{CPU (Intel)} & \textbf{CPU Speed} & \textbf{Root} & \textbf{Limitations} \\ \midrule
        1 &
           CentOS 7  & Xeon E5-2690 v3   & 2.60GHz & \xmark & Job duration \\
          & CentOS 7  & Xeon E5-2680 v4   & 2.40GHz & \xmark & \\ \hline
        2 & Ubuntu 18 & Xeon Gold 6248    & 2.50GHz & \xmark & Job count \\ \hline
        3 & Ubuntu 18 & Xeon E5-2683 v3   & 2.00GHz & \xmark & CPU limits \\ \hline
    \end{tabular}
\end{table}

We utilized three different High Performance Computing environments in order to conduct our experiments.
HPC resources are not often used for fuzzing experiments, but they provide substantial resources
if the experiments are compatible with the restrictions of the computing environment.
For this reason, we chose not to evaluate fuzzers that required root permissions or a modified kernel to execute correctly.
The specifications of the HPCs and their restrictions are shown in Table~\ref{table:hpcs}.

\hpc{1} was the only environment running Linux kernel version 3 which was required for QSYM but incompatible with Angora.
As this environment allowed for a significant number jobs to run in parallel, it was used for all the 24~h experiments
with the exception of Angora which ran on \hpc{2}.
Policies enforced by the operator of \hpc{3} made access to data generated in the environment challenging.
As such, it was used only for the long-running experiments when the other environments were unsuitable.

As described in Section~\ref{section:methodology:coverage}, we ran a monitor process concurrent with each fuzzer to track
coverage and bugs discovered over time.
In all HPC environments, Linux cgroups were used to limit the fuzzer and monitor to the necessary number of CPUs.
To ensure the monitor did not affect the fuzzer performance, two additional CPUs were allocated for the monitor process which the fuzzer was configured not to use.

\section{Evaluating Synthetic Bug Injection}
\label{section:evaluation}

We conducted eight experiments totaling approximately 733K CPU-hours. 
The merged coverage database reports the coverage for approximately 60M test case inputs which correspond to 226K unique edges from the 16 challenges.
7.3 million test case inputs triggered crashes for 3,421 unique bugs.

We here report an analysis of the data collected in these experiments.  With this analysis, our goal is to answer several research questions:
\begin{inparaenum}[\itshape (i)\upshape]
    \item What can synthetic bugs tell us about relative performance between popular fuzzers?
    \item Do different bug injection techniques yield bugs of differing difficulty to discover?
    \item How difficult are synthetic bugs to discover compared to organic bugs?
\end{inparaenum}

\subsection{Evaluation Metrics}
\label{section:setup:metrics}

\begin{figure}
    \includegraphics[width=0.85\columnwidth]{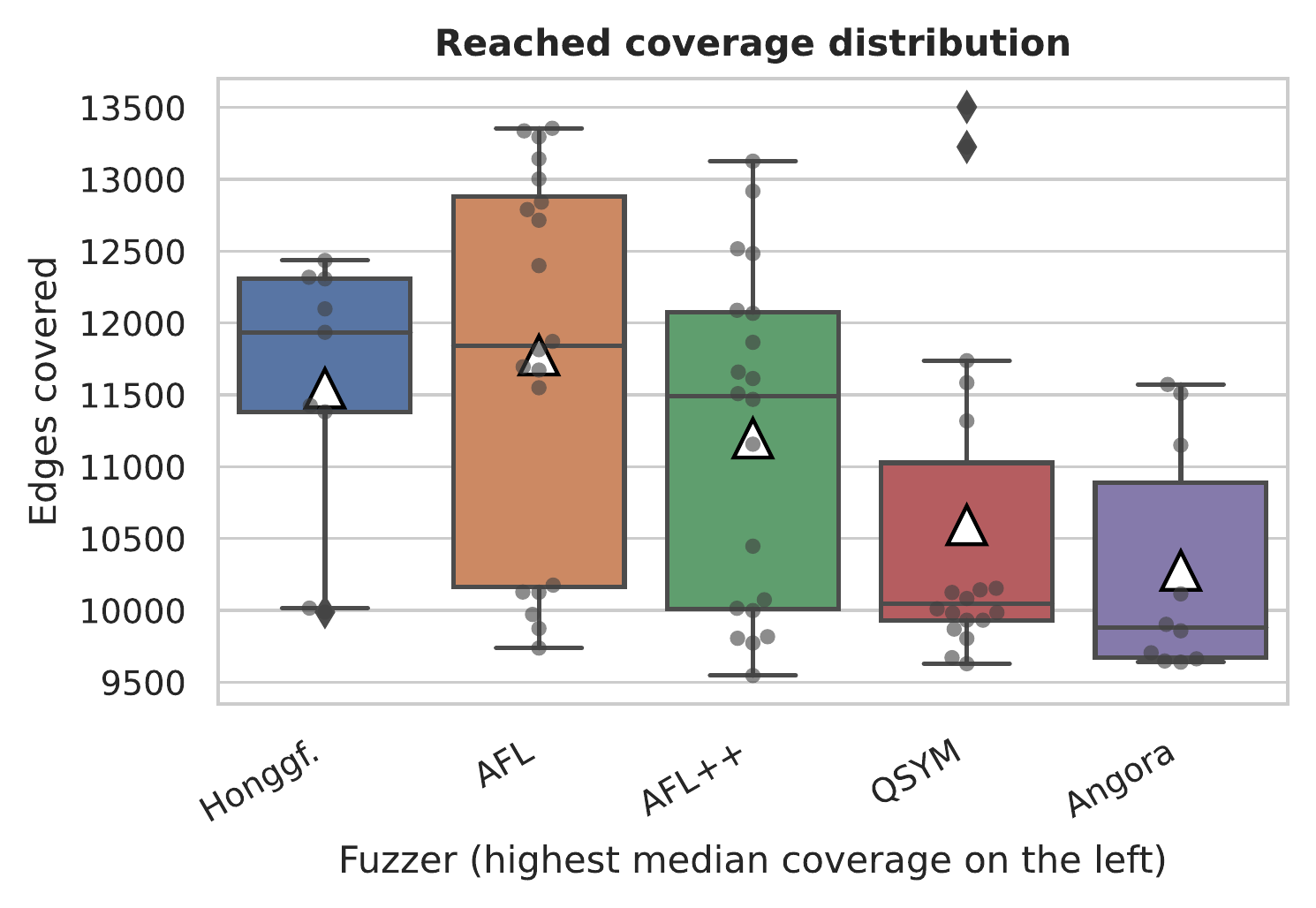}
    \caption[Example rankings]{Example top five rankings for \sqliteB.}
    \footnotesize{The white triangle in all box and whisker plots represents the mean.}
    \label{figure:stats-ranking-sqliteB}
    \Description{}
\end{figure}

Before presenting the results of the evaluation, we consider and justify the choice of evaluation metrics used herein.
A topic first addressed by Klees et al.~\cite{klees_2018_evaluatingfuzztesting}, many subsequent fuzzing studies have attempted to follow the Arcuri and Briand~\cite{arcuri2014, arcuri2011} recommendations regarding statistical tests for assessing randomized algorithms.
Despite these guidelines, reporting sound statistical measures on fuzzing experiments is a challenging task.
The data often contains outliers and the variance of the distributions is unknown, differing significantly between fuzzer-target combinations.
Unfortunately, many papers still report performance metrics like unique crashes and unique paths which can be wildly misleading.
Data regarding bugs found is overwhelmingly sparse since many fuzzers may not find any bugs and the total number found normally remains in the single digits.

\emph{Case Study: Measure of centrality.}
The amount of variance present in most experimental data makes an arbitrary choice of the measure of centrality semi-dangerous.
Choosing to report median vs.\ mean or vice-versa could change reported rankings and/or percent improvement by a non-trivial amount.
The \sqliteB challenge represents this data analysis dilemma well, shown in Fig.~\ref{figure:stats-ranking-sqliteB}.
If medians are chosen to represent data, Honggfuzz attained the most edge coverage; however, AFL takes the top spot when means are chosen.
Additionally, AFL++ falls from a 14\% increase in coverage over QSYM to 5.7\% increase using median vs.~mean respectively. 
The presence of outliers combined with highly dispersed distributions makes both the mean \emph{and} median a poor summary representation of the data and underlying distributions.

\emph{Our Evaluation Metrics.}
In Table~\ref{table:fuzzer_performance}, we report the Vargha and Delaney $\vda$ measure to rank results combined with the Mann-Whitney U test (\emph{using the exact method to determine the distribution}) to provide a statistical test of the null hypothesis that the distributions are equal.
 The $\vda$ measure provides an intuitive value that, given fuzzer $F_1$ and fuzzer $F_2$, quantifies stochastic dominance, or the probability that fuzzer $F_1$ will perform better than $F_2$.
 An $\vda$ value of 0.95 would indicate that in 95\% of the experiments $F_1$ will perform better than $F_2$, or alternatively, that in the next experiment, $F_1$ is 95\% likely to outperform fuzzer $F_2$.
We furthermore use R's Mann-Whitney implementation which is considered robust as opposed to the SciPy implementation which is \emph{badly broken}~\cite{scipyissue11035}.\footnote{SciPy's implementation is incorrect for $n<20$ as the normal approximation that it relies to compute the Mann-Whitney U test on is invalid in these cases. We note that the fuzzing community's reliance on a broken statistical test implementation for many of its evaluations has implications beyond this work.}

\subsection{Fuzzer Performance on Synthetic Bugs}
\label{section:evaluation:relative}

\begin{table}
\small
\centering
\setlength\tabcolsep{3.5pt} 
  \caption{Fuzzer performance summary.}
  \label{table:fuzzer_performance}
  \begin{tabular}{l | lrl | lrl} \toprule
                         & \multicolumn{3}{c}{\textbf{Edge Coverage}}                &  \multicolumn{3}{c}{\textbf{Bugs Found}}              \\
      \textbf{Challenge} &  \textbf{First}  &   \textbf{$A_{12}$} & \textbf{Second}  &  \textbf{First} &    \textbf{$A_{12}$}    & \textbf{Second} \\ \midrule
        duktape &            AFL-rb & \textbf{\cellcolor{green!50}0.850} &  AFL++ &              QSYM &         \cellcolor{green!10}0.611  &        AFL\\
         fileB3 &     \textbf{QSYM} & \textbf{\cellcolor{green!50}0.950} &  AFL++ &     \textbf{QSYM} & \textbf{\cellcolor{green!50}0.950} &     AFL-rb\\
         fileS3 &     \textbf{QSYM} &         \cellcolor{green!10}0.658  & Angora &     \textbf{QSYM} & \textbf{\cellcolor{green!50}0.753} & Angora\\
         fileS4 &   \textbf{Angora} &         \cellcolor{green!10}0.616  &   QSYM &   \textbf{Angora} & \textbf{\cellcolor{green!50}0.768} &     QSYM\\
         grepB2 &             AFL++ & \textbf{\cellcolor{green!50}0.876} &    AFL &          Eclipser & \textbf{\cellcolor{green!50}0.842} &     AFL-rb\\
         jpegS3 &              QSYM & \textbf{\cellcolor{green!50}0.942} &  AFL++ &          Eclipser &         \cellcolor{green!05}0.528  &       QSYM\\
            jqB &     \textbf{QSYM} &         \cellcolor{green!05}0.567  & Angora &     \textbf{QSYM} & \textbf{\cellcolor{green!50}0.886} &     Angora\\
           jqB2 &     \textbf{QSYM} & \textbf{\cellcolor{green!50}0.903} &  AFL++ &     \textbf{QSYM} & \textbf{\cellcolor{green!50}0.926} &     AFL++\\
           jqS3 &     \textbf{QSYM} & \textbf{\cellcolor{green!50}0.803} &  AFL++ &     \textbf{QSYM} & \textbf{\cellcolor{green!50}0.947} &     AFL++\\
           jqS4 &     \textbf{QSYM} &         \cellcolor{green!25}0.735  & AFL-rb &     \textbf{QSYM} & \textbf{\cellcolor{green!50}0.950} &  Honggf.\\
       newgrepS &    \textbf{Ankou} &         \cellcolor{green!50}0.822  & Honggf. &   \textbf{Ankou} &         \cellcolor{green!10}0.578  &  Honggf.\\
          pcreB &     \textbf{QSYM} & \textbf{\cellcolor{green!25}0.722} &     AFL &    \textbf{QSYM} & \textbf{\cellcolor{green!50}0.891} & Eclipser\\
        sqliteB &               AFL &         \cellcolor{green!10}0.589  & Honggf. &             QSYM & \textbf{\cellcolor{green!50}0.917} &  Honggf.\\
       tcpdumpB &              QSYM &         \cellcolor{green!05}0.541  &  AFL-rb &           Angora & \textbf{\cellcolor{green!50}1.000} & Eclipser\\
      tinyexprB2 & \textbf{Honggf.} & \textbf{\cellcolor{green!50}0.978} &  Angora & \textbf{Honggf.} & \textbf{\cellcolor{green!50}0.850} & Angora\\
         yamlB2 &            Angora & \textbf{\cellcolor{green!50}0.800} & Honggf. &             QSYM &         \cellcolor{green!10}0.610  &    AFL-rb\\ \bottomrule
         \multicolumn{7}{l}{\footnotesize $A_{12}$ values in bold indicate distributions are not equal (p-value < 0.05).}\\
         \multicolumn{7}{l}{\footnotesize Fuzzer names in bold indicate fuzzer first in both coverage and bug finding.}\\
  \end{tabular}
\end{table}

\begin{figure}
    \includegraphics[width=\columnwidth]{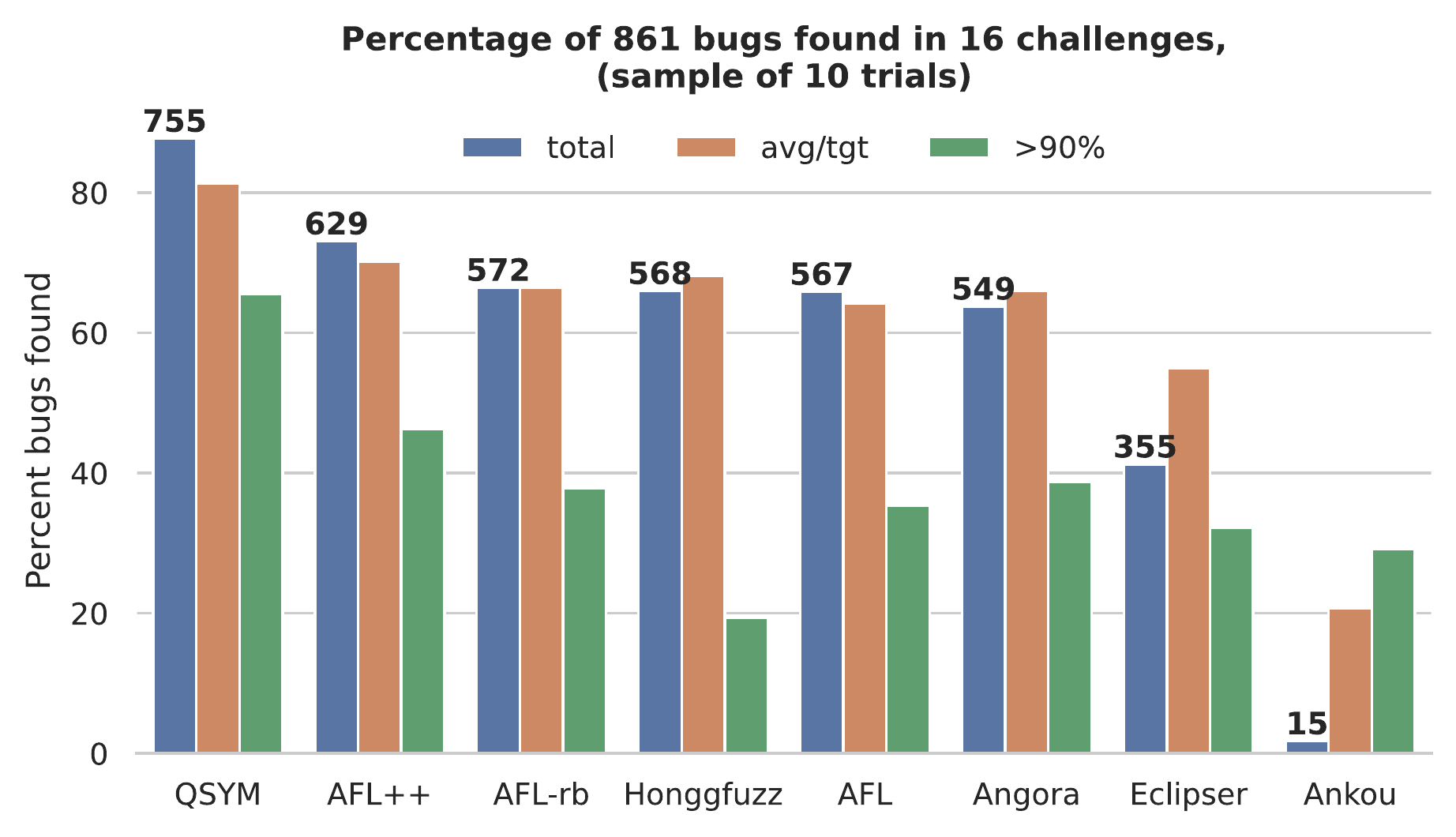}
    \caption{Overall bug finding stats.
        {\mdseries
        Results are taken from a representative selection of 10 experiments for each fuzzer.
        \emph{total} is the percentage of the total bugs found.
        \emph{avg/tgt} is the average of the percent bugs found per target.
        \emph{>90\%} is a measure of consistency; the percentage of bugs found in 9/10 experiments.
        The number above each \emph{total} bar is the raw number of bugs found by each fuzzer.}}
    \label{figure:percent_bugs_found}
    \Description{}
\end{figure}

We first turn to the question of how fuzzers perform on the synthetic bug challenge set, and whether they have utility in distinguishing the strengths and weaknesses of fuzzers under test.  
Fig.~\ref{figure:percent_bugs_found} presents an overview of relative fuzzer performance on the challenge set.  
From this, one can glean some larger trends.  QSYM clearly outperforms the field overall in absolute numbers of bugs found as well as average bugs per target.  
Other fuzzers, such as AFL-rb, Honggfuzz, and Angora have skewed performance across the challenge set since their average bug discovery rate per target is higher than the number of bugs found.  
Finally, some fuzzers show inconsistent behavior.  Honggfuzz, for instance, shows a substantially lower number of bugs found in 90\% of experiments than the total number of bugs found.  
Inconsistency suggests that average bug-finding performance might be far away from the ``best'' possible performance reported over many repeated trials.

While these larger trends are suggestive, attaining deeper insight requires individual discussion of each fuzzer. In the following, we highlight results for several fuzzers.

\emph{AFL.} 
As previously mentioned, dictionaries can be very effective in boosting AFL's performance on a per-target basis. 
Without the aid of symbolic execution, dynamic taint analysis, or even comparison byte-splitting, AFL-based fuzzers are mostly useless at finding LAVA-injected bugs. 
This is due to the nature of how LAVA bugs are triggered which often requires matching a 32-bit magic value.
However, supplied with a simple dictionary of constants extracted from a challenge (e.g., parsed by objdump) AFL can be very effective at finding LAVA bugs. 
In the subset of 24~h experiments, AFL without a dictionary only found 15/861 bugs (2.74\%) while AFL assisted with a dictionary found \emph{655}/861 bugs (74.07\%).

\emph{QSYM.} 
In our experiments, QSYM consistently achieves the best performance, discovering the most edge coverage on eight challenges and the most bugs on 10 challenges.
QSYM's main distinguishing feature is its use of concolic execution to assist in path constraint satisfaction.  
As such, the results strongly suggest that this capability is very useful for achieving greater coverage and, in turn, discovering more bugs.

\emph{Angora.}
While Angora demonstrated remarkably fast and thorough bug finding on \fileSfour, it
failed to generate new inputs on four challenges, which makes it clear that reliance on DFSan for taint tracking makes it more brittle than other fuzzers in this evaluation.

\emph{Honggfuzz.}
Honggfuzz outperforms all other fuzzers on exactly one challenge, \tinyexprBtwo. 
Honggfuzz explores the most edges and finds three of the four bugs (the remaining bug was never found by any fuzzer).
We ascribe this to Honggfuzz's unique string mutation strategies combined with its comparison operator analysis; these give it an edge for certain types of input parsers.

\emph{Ankou.}
Ankou covers the most edges and discovered the most bugs on the \newgrepS challenge, highlighting that its \emph{distance-based} fitness function can be effective under certain circumstances.

\begin{figure}
    \includegraphics[width=0.9\columnwidth]{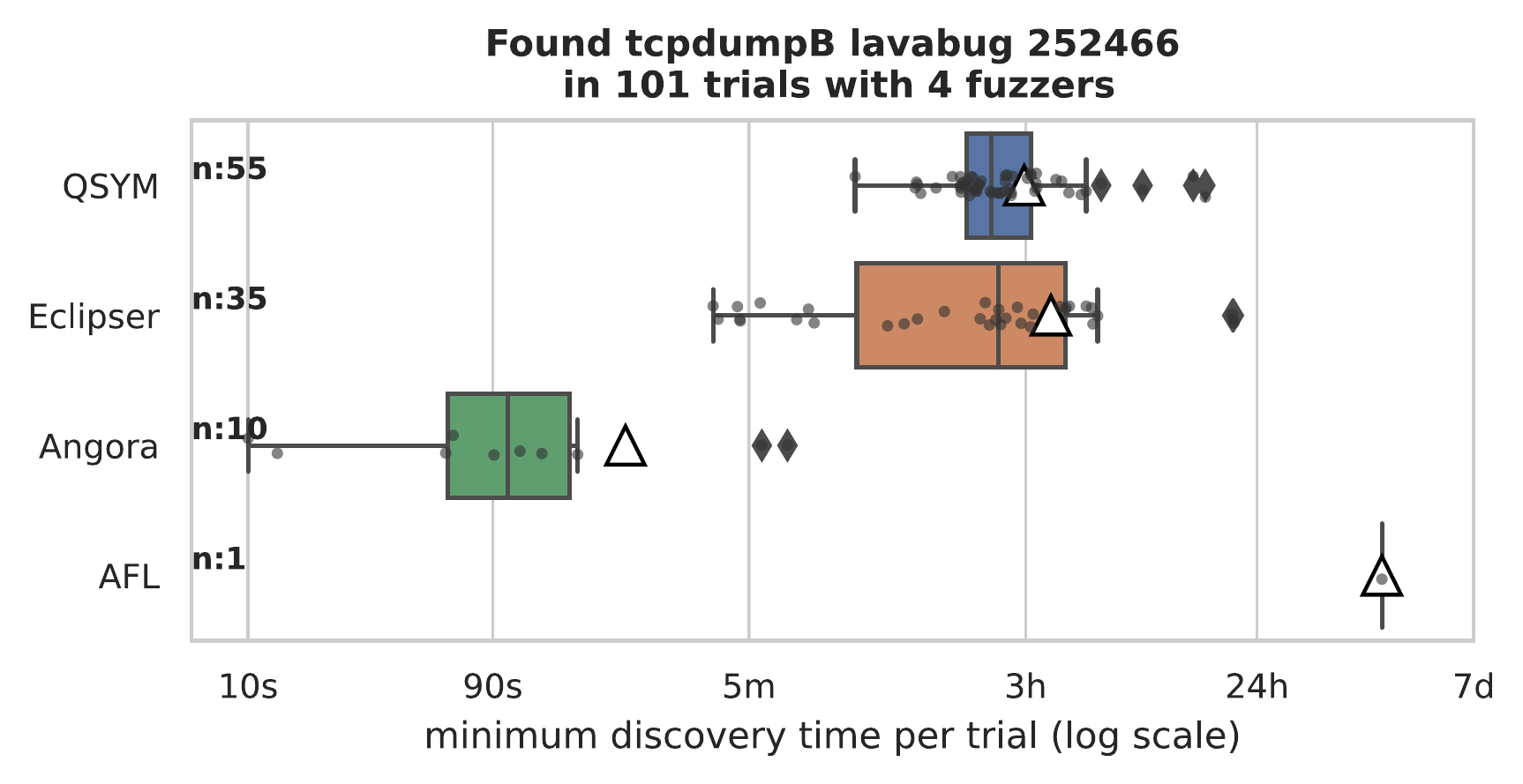}
    \caption{Find times for \tcpdumpB bug 252466.}
    {\mdseries \textbf{n:} is the number of trials where the bug was discovered}
    \label{figure:finding_bug_252466}
    \Description{boxplot of discovery times for tcpdumpB bug 252466}
\end{figure}

\emph{tcpdumpB bug 252466.}
Looking at the example LAVA bug previously shown in Fig.~\ref{figure:lavabug_252466},
we see it highlights the circumstances under which some tested fuzzers outperform the rest of the field.
The relative discovery times from all experiments for this bug are shown in Fig.~\ref{figure:finding_bug_252466}.
Angora has the best median relative discovery times of approximately 100~s.
Eclipser and QSYM are also effective and consistent at finding this particular bug, but need more time (\textasciitilde{}8,000~s) to solve the necessary constraints.
Eclipser reports the lower \emph{minimum} discovery time of 652~s while QSYM has the better \emph{median} discovery time of 7,961~s.

AFL only found this shallow bug in seven-day experiments with a median discovery time of 265,082~s, or about 3.1~d.
Multi-DUA LAVA bugs like this one renders a dictionary of constants mostly ineffective in finding LAVA bugs with AFL-based fuzzers.
Similarly, multi-DUA bugs mitigate the effectiveness of Honggfuzz's split-memory comparisons.
However, this bug is still quickly and reliably found with concolic execution or data-flow methods, since the constraints are not deeply nested in the program control flow.

\subsection{Discovery Difficulty: Synthetic Bugs}
\label{section:evaluation:lava_difficulty}

\begin{figure}
    \includegraphics[width=0.95\columnwidth]{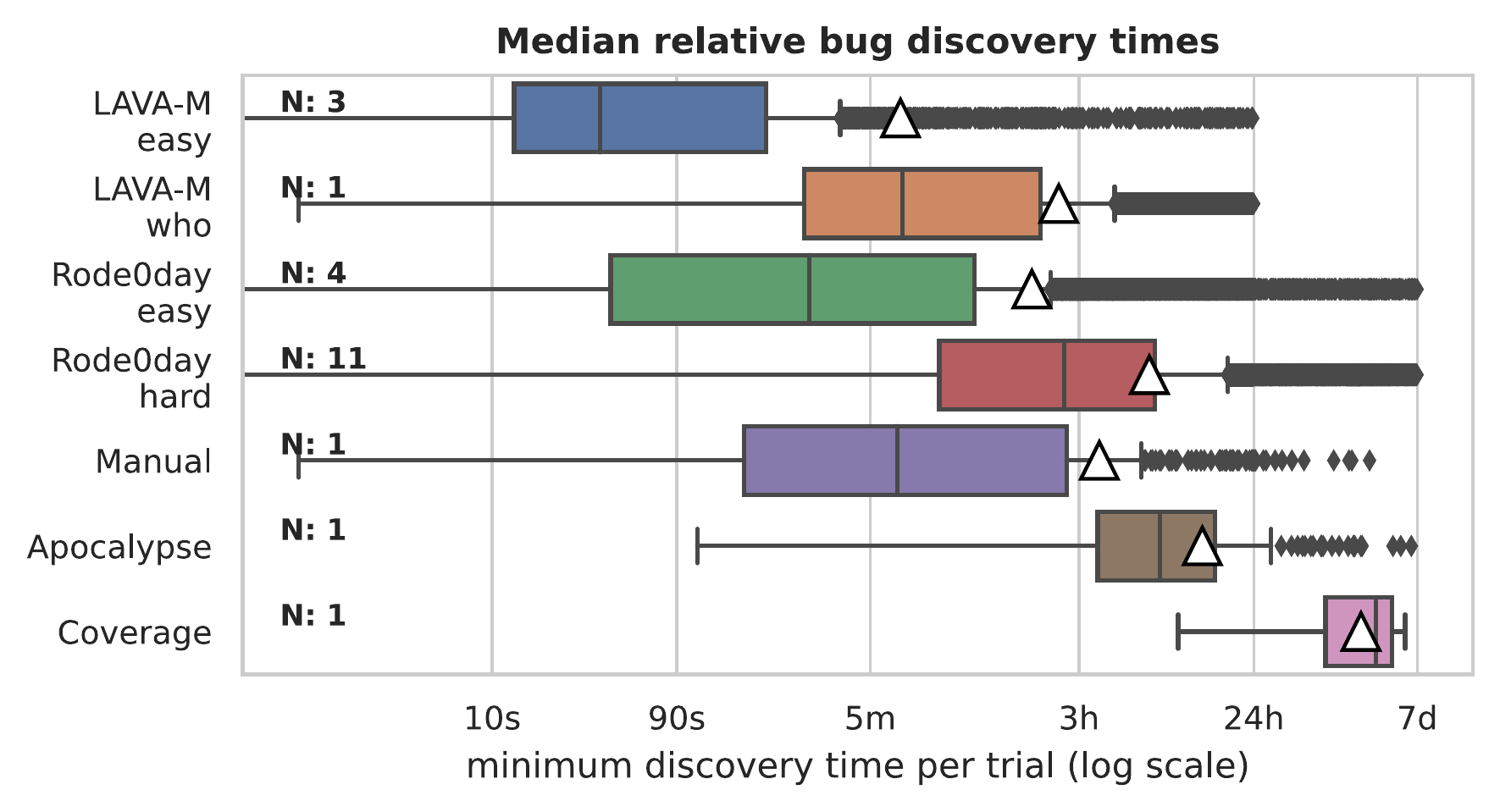}
    \caption{Discovery times of LAVA-M vs.~Rode0day bugs.\\
    {\mdseries \textbf{N:} is the number of challenges of the specified type.}}
    \label{figure:median-relative-bug-discovery}
    \Description{barplot showing discovery times of LAVA-M and Rode0day bugs}
\end{figure}

\emph{Have \Rodeoday's advancements in bug injection led to more difficult to discover synthetic bugs?} 
Our experimental results show that LAVA-M bugs are not particularly ``difficult'' to find.
We define \emph{bug difficulty} as the median time required to find a bug across all runs.
Using this metric, we find that Angora, Eclipser, and even AFL are able to find the majority of LAVA-M bugs within two hours.

Given the more sophisticated bug injection techniques used to produce the Rode0day corpus, we hypothesize that Rode0day bugs are more difficult to find. 
This does appear to be the case: while four challenges in the Rode0day corpus (\duktape, \yamlBtwo, \pcreB and \grepBtwo) approximate the LAVA-M difficulty closely, the remainder of the evaluated Rode0day challenges show increasing bug difficulty using the median bug find time metric. 
For instance, consider the various \file challenges, where only Angora is able to find the majority of injected bugs in two of the targets within the first two hours of fuzzing. 
Other fuzzers take longer to find bugs in \file challenges and fail to find all the bugs consistently.

Despite the existence of an increase in difficulty between the LAVA-M and \Rodeoday corpora, it is unclear that this represents a \emph{substantial} increase in difficulty.
In order to better understand the difference in difficulty between single- (LAVA-M) and multi-DUA (\Rodeoday) bugs, we examined several solution inputs that the fuzzers generated for bug 252466 in \tcpdumpB whose results are described above.
This multi-DUA bug is triggered when the following equation is satisfied:
$$\mathsf{ether\_dhost} \cdot \mathsf{ether\_shost} - \mathsf{ndo\_snapend} = \mathsf{0xe2d6e451}$$
The intended solution input uses three 4-byte values that are within the ASCII range of printable characters.
Several fuzzers simplify this equation by setting two of the values to identity elements (0 or 1 for addition and multiplication, respectively).
The simplified equation then becomes linear, which is substantially easier for an SMT solver to satisfy:
\begin{align*}
0 \cdot 0 - \mathsf{ndo\_snapend} &= \mathsf{0xe2d6e451}, \text{ or} \\
\mathsf{ether\_shost} \cdot 1 - 0 &= \mathsf{0xe2d6e451}
\end{align*}
This demonstrates a weakness of the current multi-DUA implementation in that the intended non-linear equation can be trivially simplified to the complexity of a single-DUA bug.
This is also likely why AFL was able to discover this bug, as the provided dictionary of constants includes the comparison value 0xe2d6e451.

Fig.~\ref{figure:median-relative-bug-discovery} shows the distributions of median bug discovery times per trial split by bug-injection technique.
Note that the x-axis is log-scale.  
The labels, ``easy'' and ``hard'' correspond to whether challenges had a median relative find time greater or less than 5~m.  
These distributions suggest that while the various categories of Rode0day bugs (LAVA, Apocalpyse, Manual, Coverage) are harder than LAVA-M bugs, the absolute increase in difficulty is not large.  
For instance, ``LAVA-M easy'' bugs can be discovered with a median time of 60~s, while ``Rode0day easy'' bugs take around 240~s.  
Even ``Rode0day hard'' bugs only require a median time of around 3~h, which is well within the computational budget of most fuzzing campaigns.
Although few challenges of other bug types are available, it appears that manually injected bugs were of similar difficulty to those added by the improved LAVA.
``Apocalypse'' bugs were more challenging to find, but only a single target was evaluated with just four bugs.
Finally, the ``Coverage'' category substantiates the claim that finding new, or previously undiscovered, coverage is exponentially difficult.

Figure~\ref{figure:median-relative-bug-discovery} highlights another important finding.  
``LAVA-M'' and the ``Rode0day easy'' challenges represent poor choices to use as benchmarks for fuzzers due to the ease and consistency that modern fuzzers discover those bugs.

\subsection{Discovery Difficulty: Rode0day vs.~Organic}
\label{section:evaluation:organic_difficulty}

\begin{figure*}[t]
    \includegraphics[width=0.9\textwidth]{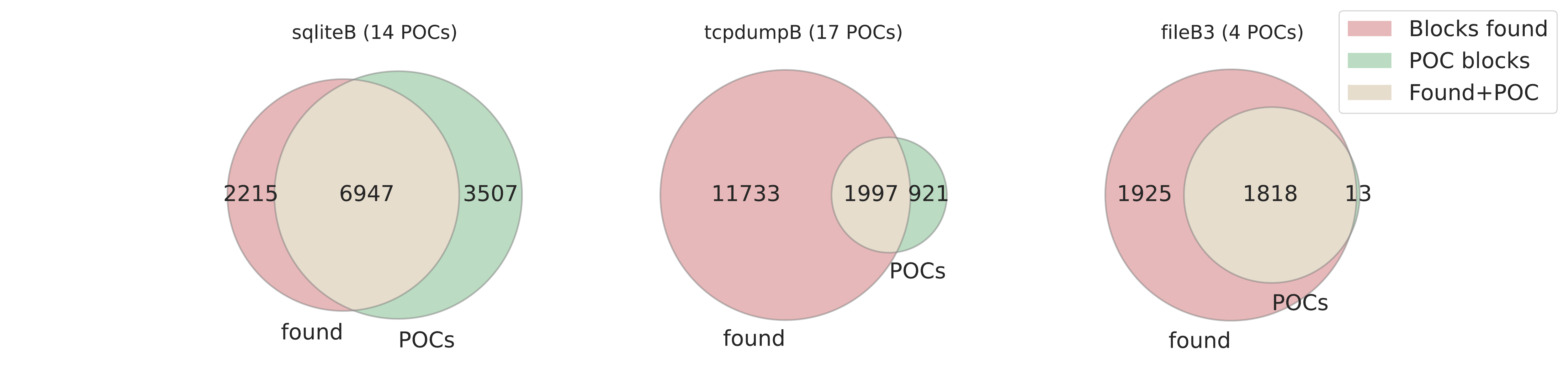}
    \caption{Blocks covered by fuzzers vs. blocks observed in POCs that trigger organic bugs.}
    \label{figure:block_coverage_venn}
    \Description{Venn diagrams comparing blocks found by fuzzers to blocks required to trigger oganic bugs}
\end{figure*}

\emph{Do Rode0day bugs approximate the discovery difficulty of organic bugs?}  This research question directly bears on the ecological validity of synthetic bugs; that is, \emph{if a fuzzer performs well on synthetic bugs, will that behavior generalize to organic bugs in real-world usage?}  One method of estimating the difficulty of finding organic bugs vs.~Rode0day bugs is to compare the median find times for organic bugs to those for Rode0day bugs.  While we cannot know \emph{a priori} the
entire set of organic bugs that are latent in the challenge corpus, we can rely on prior bug reports as a lower bound.  There are in fact at least 50 publicly reported bugs (35 with POC inputs available) that exist in the versions of the software artifacts used to create the Rode0day challenges, and thus we expected to trigger some subset of these.

Unfortunately, \emph{none} of these existing bugs were found during our experiments\footnote{In a recent \href{https://www.fuzzbench.com/reports/2021-02-17-bug-paper/index.html}{FuzzBench experiment}\cite{fuzzbench} 3 of 4 \fileBthree organic bugs were found occasionally by significantly improved versions of AFL++ and Honggfuzz; each finding a max of one bug per 24hr trial.}.  This was a counter-intuitive result; after all, most bugs are now found via fuzzing,\footnote{In fact, academic fuzzers were responsible for finding the organic bugs in \sqliteB~\cite{zhong_2020_squirreltestingdatabase} and \fileBthree~\cite{gan2020greyone}.} and the fuzzers included in our experiments are all considered (to varying degrees) state-of-the-art.  The fuzzers had also been run on the challenge corpus many times over hundreds of thousands of CPU-hours (see~Table~\ref{table:experiments}).

In investigating this phenomenon further, we uncovered several immediate reasons why some of these bugs were not found.  One such reason is that the majority of the bugs require ASan~\cite{serebryany_2012_addresssanitizerfastaddress} to be enabled in order to detect memory corruption.  Without such a ``fail-fast'' memory corruption detector in use, it is unlikely that those bugs would be rediscovered, or that if they were triggered that they would manifest in a recognizable way such that they could be linked to the original report.  Another reason is that of the 17 bugs that exist in tcpdump~v4.9.2, all require different command line arguments than those used in the Rode0day competitions.  Without the presence of those arguments, the vulnerable code would never be executed regardless of the fuzzing inputs.

We also conducted a coverage analysis of 35 organic bugs to determine if the basic blocks related to triggering them were discovered during fuzzing. To do so, we first collected proof-of-concept (POC) inputs corresponding to each bug.  Then, we verified whether or not the POCs would trigger the same fault in the Rode0day version of the program.  Finally, we used the same coverage binaries from our experiments and recorded the basic blocks covered by all the POC inputs.  Comparing the coverage achieved by the fuzzers in our evaluation to the coverage required to trigger the bugs using the POCs would provide a measure of how ``close'' the fuzzers had come to rediscovering the organic bugs.

Fig.~\ref{figure:block_coverage_venn} shows Venn diagrams of the basic blocks found during our experiments vs.~the basic blocks covered by the POC inputs for the challenges containing known bugs: \sqliteB, \tcpdumpB, and \fileBthree.  Interestingly, each of these challenges represent a range of ``closeness'' to triggering known bugs.  Using proportion of block coverage as a proxy metric for distance, we find that fully 3,507 (34\%) of the 10,454 blocks covered by POCs for known bugs in \sqliteB were \emph{not} covered by a fuzzer in any experiment we conducted.  This suggests that, on average, fuzzers did not come close to rediscovering the 14 known bugs in \sqliteB.  Moreover, given the exponential cost of covering new code~\cite{bohme_2020_fuzzingexponentialcost}, it is unlikely that the fuzzers would have found all of these bugs even given considerably more time.

\fileBthree represents the other end of the spectrum.  In this case, the fuzzers were unable to cover only \emph{13} (0.71\%) of the 1,831 blocks covered by four POCs, two of which did not require ASan to trigger.  This suggests that satisfying the path constraints to reach those last 13 blocks was exceedingly difficult for the fuzzers we tested.  It is possible that these two discoverable bugs were gated behind ``hard'' path constraints: since \file has been continuously fuzzed as part of OSS-Fuzz~\cite{google_2020_ossfuzz} for years, it can be expected that most of the low-hanging fruit, or easy-to-discover bugs, has already been picked, leaving only those bugs that are more difficult to reach.  This example certainly indicates that not all basic blocks are equally difficult to cover, and that path constraint difficulty plays a significant role in achieving coverage.  Thus, it also gives further empirical support for a power-law distribution of block and edge coverage.

\subsection{Bug Injection and the ``Main Path''}
\label{section:evaluation:main_path}

\begin{figure}[t]
    \includegraphics[width=0.9\columnwidth]{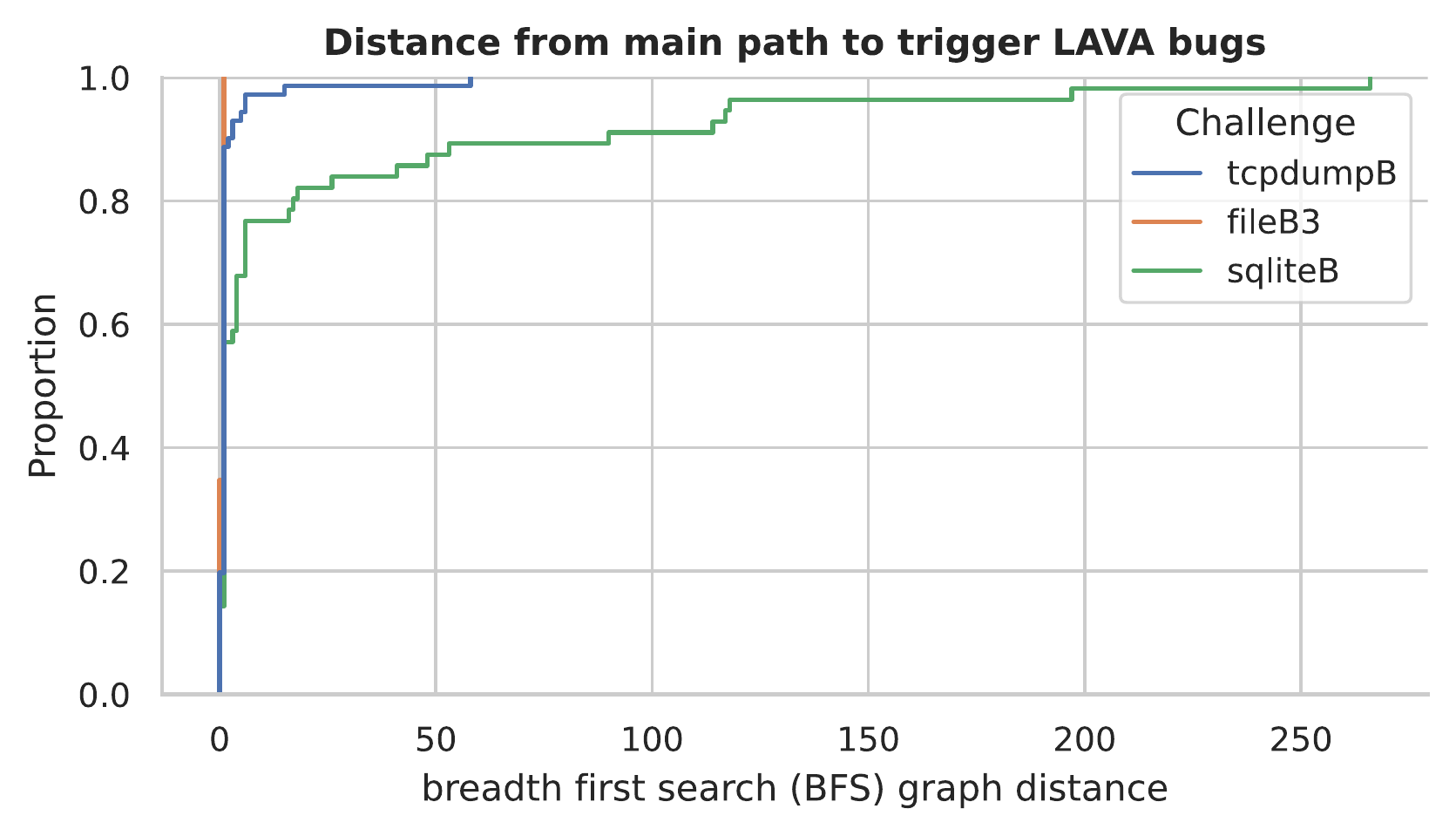}
    \caption{CDF of distance as the shortest-path number of edges between the \emph{main path} and injected bugs.
    {\mdseries
        The main path is defined as those edges with median discovery times $<1$~h in each challenge across all fuzzers and experiments.
    }}
    \label{figure:lava_bug_main_path_distance_dual}
    \Description{CDF of injected bug distance from main path}
\end{figure}

Despite the small number of organic bugs in relation to the LAVA-M and Rode0day data sets, the data strongly suggests that organic bugs are strictly more difficult to discover for state-of-the-art fuzzers.  To better understand this phenomenon, we examined the placement of injected bugs vs.~organic bugs in the challenge programs.

Due to its dependency on dynamic taint analysis, LAVA only inserts bugs along an execution path provided by a concrete input.  Since the bugs themselves are only separated from covered code by a triggering predicate that can be as simple as testing for equality with a magic value, LAVA bugs are very close to covered code and thus the difficulty of discovering them scales linearly with the computational resources invested~\cite{bohme_2020_fuzzingexponentialcost}.

Fig.~\ref{figure:lava_bug_main_path_distance_dual}~ demonstrates this phenomenon well.  This graph plots a CDF of shortest path distances between the \emph{main path} of the evaluation challenges containing organic bugs (\sqliteB, \tcpdumpB, and \fileBthree) and LAVA bugs.  Here, we define the ``main path'' as those edges with a median discovery time $<1$~h in each challenge across all fuzzers and experiments.  The main path threshold was empirically chosen to represent a common inflection point in new coverage over time from linear to logarithmic behavior---or, alternatively, the approximate transition from linear to exponential difficulty in attaining new coverage.  The shortest path distance represents a lower bound for how many edges a fuzzer would need to cover in order to trigger a bug.

For all three challenges, 85\%, or 172 of all 204 injected bugs were $\leq 1$ edge away from the respective main path.  That is, most LAVA bugs are indeed empirically very close to covered code.  The remaining 15\% of bugs injected into \sqliteB tail off to a maximum distance of $\approx 200$ edges, which coincides with the large number of POC blocks that were never covered by fuzzers as shown in Fig.~\ref{figure:block_coverage_venn}.  On the other end of the spectrum, 100\% of LAVA bugs in \fileBthree were merely $\leq 1$ edge away from easily covered code.

\begin{figure}
    \includegraphics[width=0.8\columnwidth]{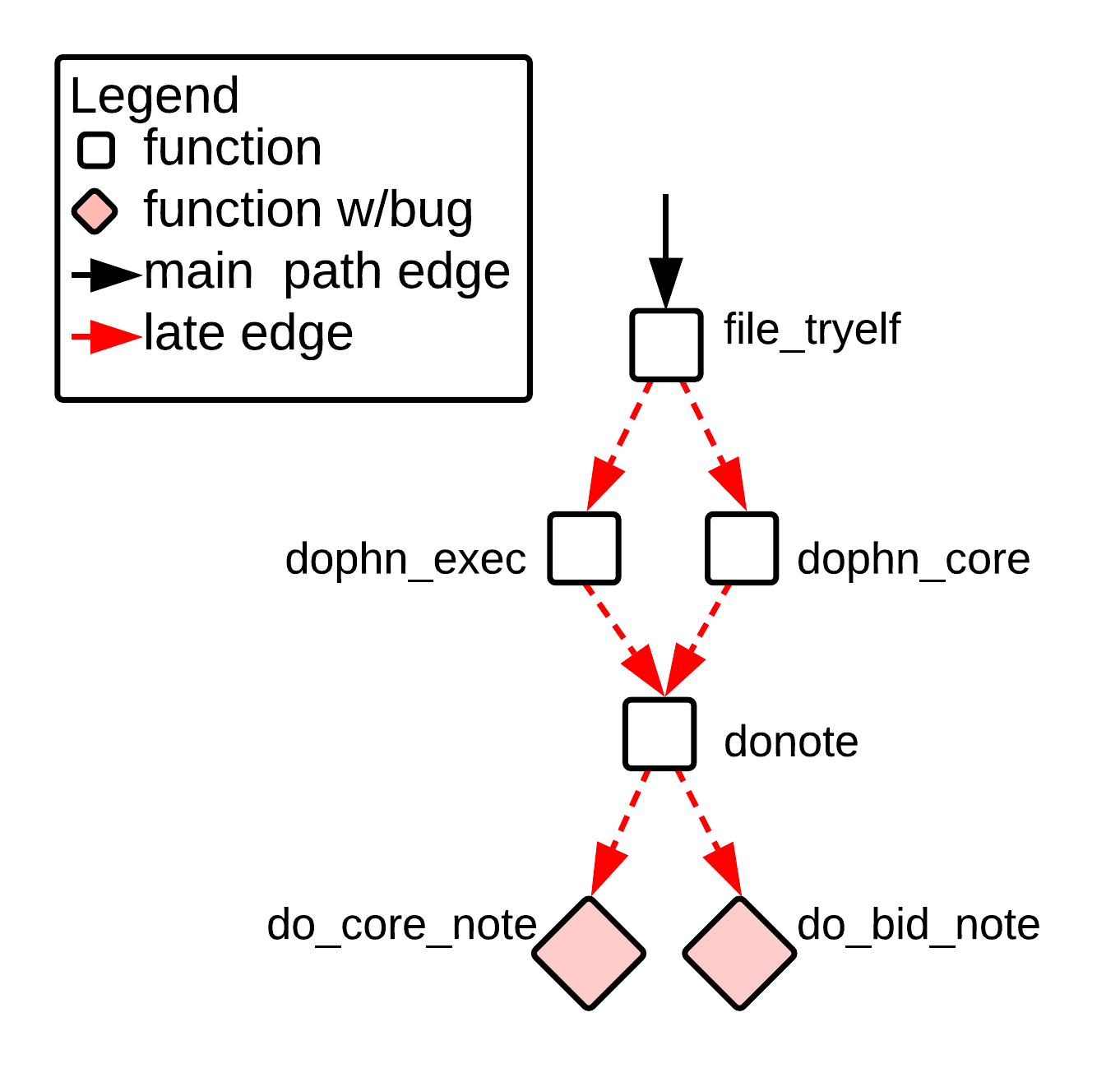}
    \caption{Partial call-graph of \fileBthree}
    \label{figure:fileB3-partial-call-graph}
    \Description{}
\end{figure}

Contrast this with the two known organic bugs in \fileBthree mentioned in \S\ref{section:evaluation:organic_difficulty}.  A manual analysis of these two bugs revealed that the shortest path between the main path and their respective trigger points required fuzzers to traverse at least two functions that were not on the main path~(see~Fig~\ref{figure:fileB3-partial-call-graph}).  This is despite the fact that only 13 POC blocks were never covered by any fuzzer, and provides a more nuanced explanation for why fuzzers were unable to discover these bugs.  That is, distance from easily covered code is characteristic of organic bugs,\footnote{We note that this can be explained in part by the challenge's significant exposure to security testing.  This is, however, a common situation for security-relevant software today.} and synthetic bugs should more closely match this distance in order to accurately model (contemporary) organic bug discovery difficulty.

\subsection{Key Findings}
\label{section:evaluation:summary}

To summarize the key findings of our synthetic bug evaluation:

\begin{enumerate}
    \item QSYM outperforms all other fuzzers on the \Rodeoday evaluation corpus.  This is likely due to its use of path constraint solving leading to higher program coverage.~(\S\ref{section:evaluation:relative})
    \item Dictionaries can \emph{drastically} affect how well AFL-based fuzzers perform, and their use (or absence) should be reported in future fuzzing experiments.~(\S\ref{section:evaluation:relative})
    \item Some fuzzers can produce near-best bug discovery performance, but are inconsistent in doing so.~(\S\ref{section:evaluation:relative})
    \item \Rodeoday bugs are not substantially harder to discover than LAVA-M bugs, suggesting that further work is necessary to generate synthetic bug challenges with better discriminatory power for future fuzzing evaluations.~(\S\ref{section:evaluation:lava_difficulty})
    \item \Rodeoday bugs are likely to be much easier to find than organic bugs, especially in programs that have been exposed to extensive security testing.  We posit that organic bugs in these programs tend to be ``far away'' from fuzzing seed paths, leading to an exponential cost in bug discovery time.  Since synthetic bugs are currently injected close to fuzzing seed paths, their discovery time remains linear.~(\S\ref{section:evaluation:organic_difficulty}-\ref{section:evaluation:main_path})
\end{enumerate}

\section{Future Directions for Bug Injection}
\label{section:future}

A major finding of our evaluation is that synthetic bugs are significantly easier to discover than organic bugs when using median discovery time as a metric.  A possible conclusion from this is that synthetic bugs are fundamentally unsuitable for fuzzing evaluations.  However, we believe that this would be a simplistic conclusion to adopt.  There are clear and substantial benefits to synthetic bug corpora in enabling scalable benchmark generation and low-cost comparative evaluation.  The community's swift adoption of LAVA-M also serves as empirical evidence of the utility of synthetic bugs.  Thus, rather than recommending that the concept of synthetic bugs be discarded wholesale, we instead consider how could they be improved to more accurately reveal real-world fuzzer performance.

\emph{Modeling organic bugs.}
One immediate direction for improving the discriminatory and predictive power of synthetic bugs is to more closely model them on organic bugs.  Recall for example that the initial version of LAVA bugs manifests as simple equality comparisons against data stored in a global array using a simple helper function (\lavaset).  In contrast, organic bugs usually do not have such a direct dependence on literal values that appear in one specific contiguous set of bytes in an input.  Rather, organic bugs tend to be triggered by conditions derived from non-trivial computation on multiple, non-contiguous bytes from an input.  Similarly, while early LAVA bugs relied on data flow to a global variable, inputs leading to the triggering of organic bugs often do not have such a simple, direct flow.  Both of these unique characteristics of LAVA bugs have been addressed to a degree in subsequent refinements of the technique (i.e., data-flow and multi-DUA injection).

However, these refinements were used in several of the evaluated programs and our results show that they still fail to sufficiently reflect the complexity of real bug data flows and input dependence.
For instance, the finding that SMT solvers can trivially convert non-linear path constraints arising from multi-DUA injection to simple linear constraints~(\S\ref{section:evaluation:lava_difficulty}) implies that more complex constraints would better match organic bug path constraints.
Hence, a potential direction to bridge this gap would be to examine the path constraints and data flows associated with organic bugs, and develop techniques to closely model the complexity of those constraints and data flows when injecting synthetic bugs.

\emph{Diversifying injection points.}
The current state of the art in synthetic bug injection is rooted in dynamic analysis.
As such, synthetic bugs are injected along a path that the injector knows how to reach.
While this technique is effective in ensuring that injected bugs can be triggered, it also biases injection towards bugs that are close to code that is ``easy'' to cover~(\S\ref{section:evaluation:main_path}).  There is building consensus that bug discovery becomes exponentially more difficult the farther away those bugs are from code that has already been covered~\cite{bohme_2020_fuzzingexponentialcost}.  Thus, the differing distance distributions from covered code between organic and synthetic bugs has a substantial impact on relative difficulty and, in turn, the predictive power of synthetic bug injection.

In our view, addressing this injection point bias is important for the viability of synthetic bug injection but also represents a major intellectual challenge.  Simply selecting arbitrary injection points does not solve the problem, as this does not guarantee that injected bugs can be triggered.  Potential solutions might involve static analyses or concolic execution to more evenly distribute synthetic bugs across a program without requiring (close) seed inputs.  However, since these program analysis techniques are also used in various ways by fuzzing tools, this inherently couples bug discovery difficulty to the capabilities of fuzzers under test.  Thus, there remains the risk that these bugs would nevertheless inaccurately model organic bug distributions and furthermore would not expose weak points of tested fuzzers.  Despite this risk, we believe that static or concolic bug injection is a worthwhile research direction.

\emph{Resisting dictionary and comparison splitting.}
As our evaluation demonstrated, dictionaries and comparison splitting are extremely effective techniques for boosting the performance of baseline mutational fuzzers like AFL and derivatives.  If synthetic bugs make use of injection techniques that are susceptible to these optimizations, then this again might render synthetic bugs easier to find and thus less indicative of true fuzzer performance.  Thus, future bug injection techniques should ensure that trigger conditions cannot be trivially satisfied from per-target dictionaries or split such that satisfying the trigger becomes logarithmic in the size of the value domain.

\emph{Quantifying the limits of hybrid fuzzers.}
Hybrid fuzzers such as QSYM that integrate concolic execution to solve path constraints clearly outperform approaches that adopt a brute-force strategy.  So long as constraint solving costs are kept in check, hybrid fuzzing turns out to be an overall win.  What is comparatively less understood is where these approaches fail.  For instance, are there particular value domains or forms that constraints take that degrade the effectiveness of constraint solvers?  As one example in this vein, recall the case of \fileBthree where fuzzers were unable to solve path constraints involving \texttt{strncmp} in order to trigger two organic bugs.  We believe that synthetic bugs could be used to systematically explore program features and constructs that hybrid fuzzers struggle with, providing valuable nuance to the analysis of hybrid fuzzers and directions for future innovation.

\section{Conclusions}
\label{section:conclusions}

In this paper, we presented a methodology for evaluating the efficacy of synthetic bug injection for comparative fuzzer evaluations.  
Using this methodology, we conducted extensive experiments totaling over 733K~CPU-hours, 36M~test cases, 23M~unique crashes, and 13~GB of block and edge coverage data from 16 challenge programs and eight fuzzers.  
Our findings show that while synthetic bugs do not approximate the difficulty of recently discovered bugs, they can provide useful insights into the strengths and weaknesses of different state-of-the-art fuzzing approaches.  
We also point out several pitfalls of conducting fair fuzzing evaluations that have not been reported in the literature.

We find that synthetic bugs are quantitatively easier to find than organic bugs using a median discovery time metric.  
Since synthetic bugs do have substantial utility and distinct scalability advantages over organic bug benchmarks, we highlight several complementary directions for improving future synthetic bug injection techniques: modeling organic bugs, injection point diversification, dictionary and comparison splitting resistance, and quantifying the limits of hybrid fuzzers.

\begin{acks}
The authors would like to thank
Northeastern University's Research Computing team,
the MIT SuperCloud and Lincoln Laboratory Supercomputing Center
and the Information Directorate's AFRL/RITB High Performance Systems Branch for
providing HPC resources that contributed to the research reported within this paper.

\noindent This material is based upon work supported by the National Science Foundation under Grant No. CNS-1916398.  Any opinions, findings, and conclusions or recommendations expressed in this material are those of the author(s) and do not necessarily reflect the views of the National Science Foundation.

\noindent This material is based upon work supported by the Office of Naval Research under Grant No. N00014-19-1-2364.  Any opinions, findings, and conclusions or recommendations expressed in this material are those of the author(s) and do not necessarily reflect the views of the Office of Naval Research.
\end{acks}

\bibliographystyle{ACM-Reference-Format}
\balance
\bibliography{bib/extracted}

\pagebreak 
\appendix

\section{Experiment Exclusions}
    \label{appendix:exclusion}

As shown in Table~\ref{table:appendix:fuzzers_used}, not every fuzzer was able to run for every experiment.
Although Angora was run on many experiments, it required building target binaries as 64-bit which sets it apart from the other fuzzers.

\begin{table}
\centering
\caption{Fuzzers run per experiment.} 
\label{table:appendix:fuzzers_used}
    \rowcolors{2}{gray!25}{white}
    \begin{tabular}{l|llllllll}
        Fuzzer   & e1     & e2     & e3     & e4     & e5     & e6     & e7     & e8     \\ \hline
        AFL      & \cmark & \cmark & \cmark & \cmark & \cmark & \cmark & \cmark & \cmark \\
        AFL++    & \cmark & \cmark & \cmark & \cmark & \cmark & \cmark & \cmark & \cmark \\
        AFL-rb   & \cmark & \cmark &        &        &        & \cmark &        &        \\
        Angora   &        & $\star$&        &        &        & $\star$ & $\star$ &$\star$ \\
        Ankou    &        & \cmark &        & \cmark & \cmark & \cmark &        &        \\
        Eclipser &        & \cmark &        & \cmark &        & \cmark & \cmark & \cmark \\
        Honggfuzz& \cmark & \cmark & \cmark & \cmark & \cmark & \cmark & \cmark & \cmark \\
        QSYM     & \cmark & \cmark & \cmark & \cmark & \cmark & \cmark & \cmark &        \\
        \multicolumn{9}{l}{$\star$ indicates fuzzer required (non standard) 64-bit targets} \\
    \end{tabular}
\end{table}

For the multi-day experiments, we chose to skip some challenges due to resource limitations.
For each of the skipped, two alternative versions of the same software (but with different bugs injected)
were analyzed.
The mapping of targets per fuzzing experiment are shown in Table~\ref{table:targets_fuzzed}.

\begin{table}
\centering
\caption{Targets fuzzed per experiment.}
\label{table:targets_fuzzed}
    \rowcolors{2}{gray!25}{white}
    \begin{tabular}{l|lllllll}
        Target(s)    & e1       & e2     & e3     & e4     & e5     & e6     & e7 \\ \hline
        LAVA-M (all) &        &        &        &        & \cmark &        &         \\
           duktape   & \cmark & \cmark & \cmark & \cmark &        & \cmark & \cmark  \\
            fileB3   & \cmark & \cmark & \cmark & \cmark &        & \cmark & \cmark  \\
            fileS3   & \cmark & \cmark & \cmark & \cmark &        & \cmark & \cmark  \\
            fileS4   & \cmark & \cmark & \cmark & \cmark &        &        &         \\
            grepB2   & \cmark & \cmark & \cmark & \cmark &        & \cmark & \cmark  \\
            jpegS3   & \cmark & \cmark & \cmark & \cmark &        & \cmark & \cmark  \\
               jqB   & \cmark & \cmark & \cmark & \cmark &        & \cmark & \cmark  \\
              jqB2   & \cmark & \cmark & \cmark & \cmark &        &        &         \\
              jqS3   & \cmark & \cmark & \cmark & \cmark &        &        &         \\
              jqS4   & \cmark & \cmark & \cmark & \cmark &        & \cmark & \cmark  \\
          newgrepS   & \cmark & \cmark & \cmark & \cmark &        & \cmark & \cmark  \\
             pcreB   & \cmark & \cmark & \cmark & \cmark &        & \cmark & \cmark  \\
           sqliteB   & \cmark & \cmark & \cmark & \cmark &        & \cmark & \cmark  \\
          tcpdumpB   & \cmark & \cmark & \cmark & \cmark &        & \cmark & \cmark  \\
         tinyexprB2  & \cmark & \cmark & \cmark & \cmark &        & \cmark & \cmark  \\
            yamlB2   & \cmark & \cmark & \cmark & \cmark &        & \cmark & \cmark  \\
    \end{tabular}
\end{table}

\section{Overall Fuzzer Performance}
Tables~\ref{table:fuzzer_performace_edges} and \ref{table:fuzzer_performace_bugs} show the 
top two fuzzers per challenge in terms of edges covered and bugs found. 
The ``First to Second'' columns show the summary statistics and the changes between them.
The p-value column is the Mann-Whitney U statistic, the Vargha-Delaney $\vda$ measure indicates the effect size, and finally the $\%~\Delta$ shows the percent change between median values.
The last two columns report the median number of edges (or bugs) attained by the first fuzzer for each challenge and the number of trials considered.

\begin{table*}
\centering
\caption{Fuzzer performance summary (edges)}
\label{table:fuzzer_performace_edges}
\begin{tabular}{l|l|l| rrr |rr}
\toprule
  Challenge &     First  &     Second &  \multicolumn{3}{c}{First to Second}                                   & \multicolumn{2}{c}{First} \\
            &            &            &                              p-value &                $\vda$   & $\%~\Delta$ & edges & N trials \\               
\midrule                               
    duktape &     AFL-rb &      AFL++ & \cellcolor{green!10}0.015 & \cellcolor{green!50}0.850 &   6.87\% & 16789 &   5 \\
     fileB3 &       QSYM &      AFL++ & \cellcolor{green!50}0.000 & \cellcolor{green!50}0.950 &  13.57\% &  6048 &  20 \\
     fileS3 &       QSYM &     Angora &                     0.179 & \cellcolor{green!10}0.658 &   2.54\% &  7862 &  19 \\
     fileS4 &     Angora &       QSYM &                     0.330 & \cellcolor{green!10}0.616 &   2.28\% &  7464 &  10 \\
     grepB2 &      AFL++ &        AFL & \cellcolor{green!50}0.000 & \cellcolor{green!50}0.876 &   1.37\% &  2334 &  20 \\
     jpegS3 &       QSYM &      AFL++ & \cellcolor{green!50}0.000 & \cellcolor{green!50}0.942 &  13.77\% &  4708 &  19 \\
        jqB &       QSYM &     Angora &                     0.588 & \cellcolor{green!05}0.567 &   0.10\% &  6807 &  18 \\
       jqB2 &       QSYM &      AFL++ & \cellcolor{green!50}0.000 & \cellcolor{green!50}0.903 &   2.86\% &  6934 &  18 \\
       jqS3 &       QSYM &      AFL++ & \cellcolor{green!25}0.001 & \cellcolor{green!50}0.803 &   0.65\% &  6805 &  19 \\
       jqS4 &       QSYM &     AFL-rb &                     0.118 & \cellcolor{green!25}0.735 &   0.20\% &  6987 &  20 \\
   newgrepS &      Ankou &  Honggfuzz &                     0.060 & \cellcolor{green!50}0.822 &   0.59\% &  6617 &   5 \\
      pcreB &       QSYM &        AFL & \cellcolor{green!10}0.025 & \cellcolor{green!25}0.722 &   1.54\% &  8765 &  17 \\
    sqliteB &        AFL &  Honggfuzz &                     0.472 & \cellcolor{green!10}0.589 &  -0.78\% & 11842 &  20 \\
   tcpdumpB &       QSYM &     AFL-rb &                     0.820 & \cellcolor{green!05}0.541 &   5.94\% & 26338 &  17 \\
 tinyexprB2 &  Honggfuzz &     Angora & \cellcolor{green!50}0.001 & \cellcolor{green!50}0.978 &   6.32\% &   656 &   9 \\
     yamlB2 &     Angora &  Honggfuzz & \cellcolor{green!10}0.028 & \cellcolor{green!50}0.800 &   0.93\% &  7849 &  10 \\
\bottomrule
\end{tabular}
\end{table*}

\begin{table*}
\centering
\caption{Fuzzer performance summary (bugs)}
\label{table:fuzzer_performace_bugs}
\begin{tabular}{l|l|l| rrr |rr}
\toprule
  Challenge &     First  &     Second &  \multicolumn{3}{c}{First to Second}                                   & \multicolumn{2}{c}{First} \\
            &            &            &                              p-value &                $\vda$   & $\%~\Delta$ &  bugs & N trials \\
\midrule
    duktape &       QSYM &        AFL &                      0.151 & \cellcolor{green!10}0.611 &   0.00\% &   13 &  20 \\
     fileB3 &       QSYM &     AFL-rb &  \cellcolor{green!25}0.002 & \cellcolor{green!50}0.950 &  91.67\% &   69 &  20 \\
     fileS3 &       QSYM &     Angora &  \cellcolor{green!10}0.029 & \cellcolor{green!50}0.753 &   8.41\% &  116 &  19 \\
     fileS4 &     Angora &       QSYM &  \cellcolor{green!10}0.020 & \cellcolor{green!50}0.768 &  11.18\% &   94 &  10 \\
     grepB2 &   Eclipser &     AFL-rb &  \cellcolor{green!10}0.013 & \cellcolor{green!50}0.842 &   2.94\% &   35 &  19 \\
     jpegS3 &   Eclipser &       QSYM &                      0.330 & \cellcolor{green!05}0.528 &    nan\% &    0 &  18 \\
        jqB &       QSYM &     Angora &  \cellcolor{green!50}0.001 & \cellcolor{green!50}0.886 &  68.75\% &   27 &  18 \\
       jqB2 &       QSYM &      AFL++ &  \cellcolor{green!50}0.000 & \cellcolor{green!50}0.926 &  95.19\% &  102 &  18 \\
       jqS3 &       QSYM &      AFL++ &  \cellcolor{green!50}0.000 & \cellcolor{green!50}0.947 &  91.67\% &   23 &  19 \\
       jqS4 &       QSYM &  Honggfuzz &  \cellcolor{green!50}0.000 & \cellcolor{green!50}0.950 &  50.00\% &   21 &  20 \\
   newgrepS &      Ankou &  Honggfuzz &                      0.649 & \cellcolor{green!10}0.578 &   0.00\% &    2 &   5 \\
      pcreB &       QSYM &   Eclipser &  \cellcolor{green!50}0.000 & \cellcolor{green!50}0.891 &  30.38\% &  103 &  17 \\
    sqliteB &       QSYM &  Honggfuzz &  \cellcolor{green!50}0.000 & \cellcolor{green!50}0.917 &  22.22\% &   22 &  18 \\
   tcpdumpB &     Angora &   Eclipser &  \cellcolor{green!50}0.000 & \cellcolor{green!50}1.000 & 102.86\% &   36 &  10 \\
 tinyexprB2 &  Honggfuzz &     Angora &  \cellcolor{green!25}0.002 & \cellcolor{green!50}0.850 &  50.00\% &    3 &   9 \\
     yamlB2 &       QSYM &     AFL-rb &                      0.450 & \cellcolor{green!10}0.610 &   0.00\% &   44 &  21 \\
\bottomrule
\end{tabular}
\end{table*}

\end{document}